\documentclass[structabstract]{aa}  
\usepackage{graphicx}
\usepackage{txfonts}
\usepackage{appendix}
\usepackage{natbib}
\usepackage{longtable, portland}
\usepackage{supertabular}
\usepackage{rotating}
\bibpunct{(}{)}{;}{a}{}{,}

\def\nh{{N$_{\rm H}$}}

\def\cm2{{cm$^{-2}$}}

\begin{document}
   \title{AGN-Host Galaxy Connection: Morphology and Colours of X-ray Selected AGN at z\,$\le$\,2}


   \author{M. Povi\'c\inst{1, 2, 3}\fnmsep\thanks{email: mpovic@iaa.es}, M. S\'anchez-Portal\inst{4,5}\fnmsep\thanks{email: miguel.sanchez@sciops.esa.int}, A. M. P\'erez Garc\'ia\inst{1, 6,5}, A. Bongiovanni\inst{1,6,5}, 
J. Cepa\inst{5, 1}, M. Huertas-Company\inst{7, 8}, M. A. Lara-L\'opez\inst{1, 9}, M. Fern\'andez Lorenzo\inst{1, 3}, A. Ederoclite\inst{1, 10}, E. Alfaro\inst{3}, 
H. Casta\~neda\inst{11}, J. Gallego\inst{12}, J. I. Gonz\'alez-Serrano\inst{13}, \and J. J. Gonz\'alez\inst{14}
          }

   \institute{Instituto de Astrof\'isica de Canarias (IAC), La Laguna, Tenerife, Spain
         \and
             Astrophysics and Cosmology Research Unit (ACRU), School of Mathematical Sciences, University of Kwa-Zulu Natal (UKZN), Durban, South Africa
         \and
	      Instituto de Astrof\'isica de Andaluc\'ia (IAA), Granada, Spain             
         \and
	      Herschel Science Centre (HSC), European Space Agency Centre (ESAC)/INSA, Villanueva de la Ca\~nada, Madrid, Spain
	 \and 
Asociaci\'on ASPID, Apartado de Correos 412, La Laguna, Tenerife, Spain
         \and
Departamento de Astrof\'isica, Universidad de La Laguna (ULL), La Laguna, Tenerife, Spain
         \and
GEPI, Paris-Meudon Observatory, Meudon, France
         \and
University of Paris, Paris, France
         \and
Australian Astronomical Observatory (AAO), Sidney, Australia
         \and
Centro de Estudios de F\'isica del Cosmos de Arag\'on (CEFCA), Teruel, Spain 
         \and
Escuela Superior de F\'isica y Matem\'aticas, Intituto Polit\'ecnico Nacional (IPN), Mexico D.F, Mexico
         \and
Departamento de Astrof\'isica y CC. de la Atm\'osfera, Universidad Complutense de Madrid, Madrid, Spain
         \and
Instituto de F\'isica de Cantabria, CSIC-Universidad de Cantabria, Santander, Spain
         \and
Instituto de Astronom\'ia UNAM, M\'exico D.F, M\'exico
             }


 
  \abstract
   {The connection between AGN and their host galaxies has been widely studied over recent years, showing it to be of great importance for providing answers to some fundamental questions 
    related with AGN fuelling mechanisms, their formation and evolution.}
   {Using X--ray and one of the deepest broad-band optical data sets, morphology and colours have been studied in relationship 
    with X--ray properties for sources at redshifts z\,$\le$\,2.0, using a sample of 262 AGN in the Subaru/XMM-\textit{Newton} Deep Survey (SXDS).}
   {Morphological classification has been obtained using the new galSVM code. Colour-magnitude diagrams have been studied in relationship with redshift, morphology, X--ray obscuration, 
    and X--ray-to-optical flux ratio. The significance of different regions has been analysed on colour-magnitude diagrams, relating the observed properties of AGN 
    populations with some models of their formation and evolution.}
   {Our morphological classification suggests that different mechanisms 
may be responsible for triggering the nuclear activity in galaxies. Observing populations of 
X--ray detected AGN on both colour-magnitude and colour-stellar mass diagrams, the highest number of sources is found to reside in the green valley at redshifts $\approx$\,0.5\,--\,1.5. However, a 
higher number of low luminosity AGN has been detected in comparison with previous works due to the high depth of the SXDS optical data. 
Wheather AGN are being hosted by early- or late-type galaxies, no clear relationship has been found with the optical colours (independently of redshift), as typical for normal galaxies. Both early- and late-type AGN cover 
similar ranges of X--ray obscuration, with both unobscured and obscured sources.}
   {Our findings may confirm some previous suggestions, that X--ray selected AGN residing in the green valley 
present a transition population, quenching star formation through different AGN feedback mechanisms and evolving to red sequence galaxies. They might be hosted by similar sources 
(majority of sources being late-type elliptical and lenticular galaxies, and early-type spirals), having similar stellar populations, being triggered mainly through major and/or minor mergers, 
and in some cases through some secular mechanism, as shown in previous numerical simulations. In the mentioned transition we observe different phases of AGN activity, with some AGN being in the 
'QSO-mode': detected as compact, blue, and unobscured in X--rays, and with others passing through different phases before and after the 'QSO-mode', being obscured and unobscured in X--rays, respectively.}

   \titlerunning{Morphology and Colours of AGN at Redshifts z\,$\le$\,2 in the SXDS Field}
   \authorrunning{Pov\'ic et al.}
   \maketitle

%

\section{Introduction}
\label{sec1}

Active Galactic Nuclei (AGN) play an important role in many aspects of modern cosmology, and of particular interest is the issue of the interplay between AGN and their host galaxies. 
The study of AGN host galaxies is shown to be of great importance for providing answers to some still unanswered questions. These questions include the effect of the AGN on their host galaxy and vice-versa, the 
origin of the accretion material, the triggering mechanisms that initiate the active phase in a galaxy, the duration of the 
active phase, etc., therefore putting important constraints on the models of supermassive black hole (SMBH) formation and growth, as well as the formation and evolution of galaxies.\\
\indent Over recent years, fundamental relationships between the AGN and host galaxy properties have been suggested, particularly with their bulges, showing the connection between galaxy 
formation and AGN activity. This involves the discoveries that most of the close massive galaxies have a SMBH in their centres \citep{magorrian98}, that local AGN predominantly reside in galaxies dominated by a 
massive bulge \citep{kauffman03}, and that the masses of the central SMBHs in nearby galaxies correlate with several host bulge properties, including luminosity \citep{kormendy95, mclure00, marconi03}, 
velocity dispersion \citep{ferrarese00, gebhardt00}, mass \citep{magorrian98, ferrarese04}, and galaxy light concentration \citep {graham01}. Moreover, in a wide range of redshifts it has been obtained 
that early-type galaxies, having more massive black holes in their centres, have lower Eddington ratios compared with later-type galaxies \citep{hickox09, povic09a, povic09b}.

Morphology and colours are two key elements used in order to study the properties of AGN host galaxies, their connection with AGN, and their evolution. They represent two of the most accessible indicators of the galaxy's 
physical structure, being crucial to understand the formation of galaxies throughout cosmic history and to provide answers to some of the fundamental questions mentioned above.\\
\indent The morphological study of AGN host galaxies has been one of the active research fields over the past years, leading to the inconsistency between the results obtained when using samples at 
both low and high redshifts, and different methods of morphological classification. The first important morphological study of local AGN (mostly Seyfert galaxies) showed that most of them reside 
in spiral galaxies \citep[e.g.,][]{adams77, heckman78, ho95}. However, \cite{kauffman03} analysed thousands of low-redshift (z\,$\leq$\,0.4) AGN host galaxies from the Sloan 
Digital Sky Survey (SDSS), and found that most AGN reside in massive galaxies, having distributions of sizes, stellar surfaces, mass densities, and concentrations, all similar to those of early-type 
SDSS galaxies. On the other hand, \cite{choi09} found that most AGN from the SDSS survey reside in late-type galaxies with intermediate luminosities and velocity dispersions. Deep surveys 
made it possible to study the morphological properties of intermediate- and high-redshift AGN host galaxies selected at different wavelengths (mainly X--ray, optical, IR, and radio). Most of these 
studies found that X--ray selected AGN usually reside in spheroid/bulge-dominated galaxies \citep[e.g.,][]{pierce07,povic09a}; however some studies 
found higher concentrations of later-types \citep{gabor09}. 

Beside morphology, colours are also important to for revealing the nature of AGN host galaxies. It is well known from previous studies of colour-magnitude relations that, in general, normal galaxies may be located in the 
'red sequence', populated by massive, bulge-dominated galaxies having older, passively evolving stellar populations, or in the 'blue cloud', populated by blue, star-forming galaxies of 
small and intermediate masses \citep[e.g.,][]{baldry04,weiner05, cirasuolo05}. In most (if not all) studies, galaxies hosting AGN lie predominantly in the 'green valley' of the colour-magnitude 
diagrams, a transition region located between the red sequence and the blue cloud \citep[e.g.,][]{barger03,sanchez04,nandra07,georgakakis08,silverman08,treister09}. This has been considered 
as one of the pieces of evidence of a connection between AGN and galaxy evolution, suggesting that AGN feedback mechanism may play an important role in regulating (quenching) star formation, 
moving the galaxies from blue, star-forming to passive, red sequence galaxies \citep[e.g.,][]{springel05,schawinski06,hasinger08}. Recently, \cite{hickox09} studied three populations of AGN, selected in radio, 
X--ray, and IR, and found that most radio selected AGN are hosted by red sequence galaxies, that X--ray selected AGN occupy all areas on the colour-magnitude diagram, but 
mostly the green valley, while most IR selected AGN reside in slightly bluer galaxies compared with the two previous populations.
However, other results show that there is no strong evidence in AGN host galaxies for either highly suppressed star formation, which is expected if AGN are responsible for star formation quenching, or elevated star formation, 
when compared to galaxies of similar stellar masses and redshifts \citep{alonso08,brusa09}.\\
\indent As mentioned above, although many analyses have been carried out over the past few years in order to study the morphological and colour properties of AGN host galaxies, there are 
still many inconsistencies between the results obtained and their interpretation. Particularly, the interpretation of morphology still remains a problem in the framework of galaxy evolution. 
Since the quality of the measured morphology is strongly dependent on the image resolution, the morphological classification in deep surveys is still very difficult, especially when dealing 
with faint and high redshift sources. Therefore, additional studies are necessary using deep observations for obtaining large samples of AGN, and testing new methods for morphological classification, in 
order to reveal the nature of AGN host galaxies, their connection with AGN, and to study their evolution throughout cosmic time.\\
\indent In this work we perform a study of the morphological and colour properties of a sample of 262 X--ray detected AGN at redshifts z\,$\le$\,2 in the Subaru/XMM-\textit{Newton} Deep Survey. 
\citep[SXDS;][]{furusawa08} field, using one of the deepest optical datasets available to the astronomical community. Morphological classification has been obtained using galSVM 
\citep{huertas08,huertas09}, one of the new codes especially useful for analysing morphology when dealing with low spatial resolution and high redshift data. A set of different morphological 
parameters has been studied, and an additional visual classification has been performed as well. The evolution of the AGN host galaxies on the colour-magnitude diagrams has been studied in 
four redshift intervals (up to z\,$\le $\,2 with bins of 0.5), for different morphological and X--ray types, and for objects having different X--ray-to-optical (X/O) flux ratio. Different 
regions on the colour-magnitude diagrams have been analysed, relating the observed AGN properties in each region with some current models of AGN formation and evolution. Finally, this paper provides the scientific community 
with a catalogue of a large sample of AGN detected in the SXDS field, including their photometric X--ray and optical data, morphological properties, rest-frame colours, and redshifts. The catalogue can be used in further studies related with AGN populations. This paper is part of the preparatory work for the long-term OTELO\footnote{OSIRIS Tunable Emission Line Object survey} survey \citep{cepa08}. OTELO is an on-going emission line survey using the Tunable Filters of 
OSIRIS\footnote{Optical System for Imaging and low Resolution Integrated 
Spectroscopy} at the GTC\footnote{Gran Telescopio de Canarias; http://www.gtc.iac.es/} telescope. The survey aims for perfoming a narrow-band tomography over a 21nm window centered at 920nm 
in at least two selected fields, and its science cases include the study of SFR density and chemical evolution in 
the Universe, high redshift QSO and AGN at any redshift, emission line ellipticals and Galactic emission line stars 
\citep{cepa07,povic09a,cepa11,lara11}.\\
\indent The paper is structured as follows: in Section~\ref{sec_data}, we describe the observational data used in this work, including a brief summary of the X--ray data processing and 
source detection, optical broad-band data, selection of X--ray emitters with optical counterparts, and estimation of the k-corrections and photometric redshifts. Section~\ref{sec_morph} 
describes the morphological classification of X--ray selected AGN. All analysis related with AGN colours is described in Section~\ref{sec_colours}, being considered in relation with redshift, 
morphology, X--ray obscuration, and X/O flux ratio, comparing the obtained results with some models of AGN formation and evolution. Finally, Section~\ref{sec_conclusions} summarises the 
main results obtained in this work. The concordance cosmology with $\Omega_{\Lambda}$\,=\,0.7, $\Omega_{M}$\,=\,0.3, and H$_0$\,=\,70\,km\,s$^{-1}$\,Mpc$^{-1}$ is assumed. Unless otherwise specified, all 
magnitudes are given in the AB system. 

The catalogue presenting the data obtained in this work is described in the Appendix and presented in the electronic edition.

\section{Observational material and data reduction}
\label{sec_data}
The Subaru/XMM-\textit{Newton} Deep Survey \citep[SXDS;][in prep.]{sekiguchi} is a large
survey, covering a contiguous region of $>$\,1\,square degree
centred at RA\,=\,02$^h$18$^m$ and DEC\,=\,-05$^{\circ}$00', with limiting AB magnitude at 3$\sigma$ of 28.4 in the B band, and with the typical seeing of 0.8 \citep{furusawa08}. 
The SXDS field has been observed at different wavelengths, from X--rays to radio. We
briefly describe the data sets used in this work below. X--ray and optical broad-band data \citep{povic09b,furusawa08} are the main data used in our analysis in order to detect and select the AGN 
populations, and to derive their nuclear and
morphological properties. Additionally, near- and mid-IR data have been used in order to measure k-corrections and photometric redshifts.

\subsection{X--ray data}
\label{subsec_Xdata}
The SXDS was observed by XMM-\textit{Newton} during the years 2000, 2002, and 2003
(PI Michael G. Watson). Seven pointings were obtained in the 0.2--10\,keV energy range.
The central observation is the deepest one, with a nominal exposure time of $\sim$\,100\,ksec, being surrounded by
six shallower observations, each of them with an exposure time of $\approx$\,50\,ksec. The surveyed area is $\simeq$\,1.14\,deg$^2$, covering the complete region 
of 5 mosaic, optical images taken with the Subaru telescope 
(see below).\\

\begin{figure}[ht!]
\centering
\includegraphics[width=9.0cm,angle=0]{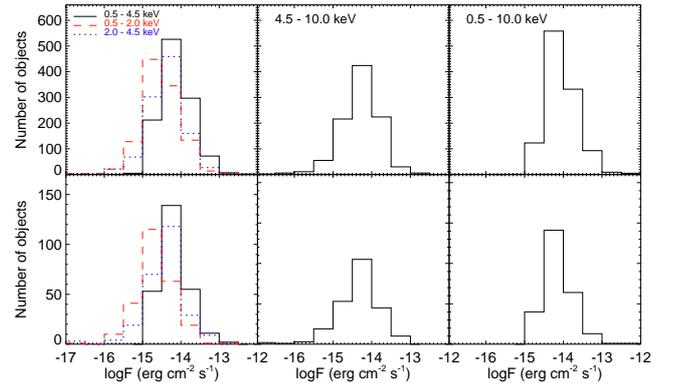}
\centering
\protect\caption[ ]{X--ray flux distributions for all 1121 X--ray detected sources (\textit{Top}) and for 262 objects with z\,$\le$\,2.0 analysed in this paper (\textit{Bottom}; 
see Section~\ref{subsec_sample_selection}). Flux distributions are represented in the 0.5\,-\,2.0 (dashed red line), 2.0\,-\,4.5 (dotted blue line), and 0.5\,-\,4.5\,keV (solid black line) energy 
bands (\textit{left panels}), 4.5\,-\,10.0\,keV (\textit{middle panels}), and 0.5\,-\,10.0\,keV range (\textit{right panels}).
\label{fig_fx_distribution}}
\end{figure}

\begin{figure}[ht!]
\centering
\includegraphics[width=8.0cm,angle=0]{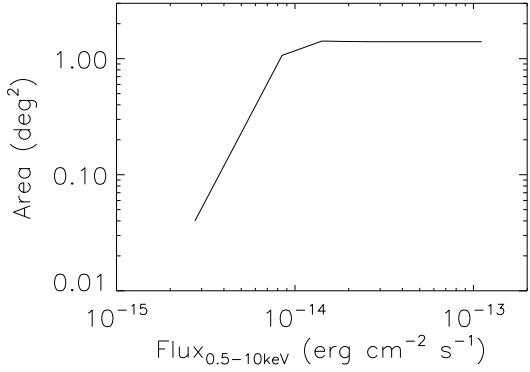}
\centering
\protect\caption[ ]{Survey area as a function of flux in the total (0.5--10keV) band.
\label{fig_skycoverage_flux}}
\end{figure}

The required data have been gathered from the XMM-\textit{Newton} v.5.0 scientific archive
(XSA\footnote{http://xmm.esac.esa.int/xsa/}). Data processing was carried out by means of the Science Analysis System (SAS\footnote{http://xmm.esac.esa.int/external/xmm\_data\_analaysis}) v7.1.2 
package, using the latest relevant Current Calibration Files (CCF). The raw Observation Data Files (ODF) were processed using the standard SAS tasks \textit{emproc} (for 
MOS cameras) and \textit{epproc} (for \textit{pn} camera) to produce calibrated event lists. Light curves were obtained by means of \textit{xmmselect} and \textit{OGIP light curve} tasks, selecting 
only events with pattern 0-4 (single and double) and PI 200-12000 for the pn camera, and pattern 0-12 and PI 200-15000 for the MOS cameras. Good Time Intervals (GTIs) were created using the \textit{tabtingen} task, 
filtering 
event lists with RATE parameter. To merge event lists and additional files we used \textit{merge} task. Six energy bands were selected, as shown in Table \ref{tab_enrang_ecf_flux}. 
Moreover, combining soft and hard bands we obtained fluxes in the 0.5\,-\,4.5\,keV energy range. This band is the most used in our analysis for comparing our results with previous ones. 
Figure~\ref{fig_fx_distribution} shows the distribution of fluxes in five energy bands, for all detected X--ray sources, and for 262 sources analysed in this paper (see Section~\ref{subsec_sample_selection}). 
The \textit{evselect} task was used 
to produce the images and \textit{fits} files in all selected energy ranges, separately for each instrument, and selecting only events with FLAG\,=\,0. We have computed the survey area as a function of the 
completeness flux in the total band (see Figure~\ref{fig_skycoverage_flux}) by evaluating the 
histogram of the survey distribution for increasing area values. The 
completeness flux for that area is derived as the energy bin 
corresponding to the maximum of the histogram. The survey is complete at 
f$_{0.5-10keV}$\,$\approx$\,3.6\,$\times$\,10$^{-15}$\,erg\,cm$^{-2}$\,s$^{-1}$ 
for the maximum survey area, $\sim$\,1.4\,deg$^2$.\\

Source detection was performed by means of the {\it edetect\_chain} SAS procedure. To minimize the number of spurious detections the likelihood threshold 
parameter\footnote{$L\,=\,-\,\ln (1-P)$, where $P$ is the probability to have a spurious detection due to random Poisson fluctuations} was set 
at L\,=\,14. Thus the probability that the source exists is at least P\,=\,0.9999916847, or less than one fake source per instrument, per pointing, and per band. In order to convert count rates to 
energy fluxes we computed energy conversion factors (ecf) using 
PIMMS (Mukai, 1993), using power law function with a spectral index $\Gamma$\,=1.8 \citep[e.g.,][]{mateos05}, and a galactic absorption nH\,=\,2.5\,$\times$\,10$^{20}$\,cm$^{-2}$ \citep{dickey90}. 
PIMMS measured all XMM-\textit{Newton} ecf values using the aperture of 15\,arcsec, and ecf values had to be multiplied by a factor of 1.47 (aperture correction) in order to obtain good 
superposition with the XMM-\textit{Newton} PSF and to integrate the flux of the source to the large distances (Saxton, R. priv. comm. and XMM-\textit{Newton} team for more information). The number of sources in 
each energy band, final ecf 
values, and median and limiting fluxes are listed in Table~\ref{tab_enrang_ecf_flux} for each energy range. 
Hardness ratios\footnote{Defined as: $HR\left(\Delta_1E/\Delta_2E\right) = \frac{CR\left(\Delta_1E\right) - CR\left(\Delta_2E\right)}{CR\left(\Delta_1E\right) + CR\left(\Delta_2E\right)}$, 
where $\Delta_1E$ y $\Delta_2E$ are different energy bands and $CR\left(\Delta_nE\right)$ is the
count rate in a given energy band} have been also defined. We have used HR(hard/soft)\,$\equiv$\,HR$(2-4.5keV/0.5-2keV)$ in this work. \\

Source detection has been performed 
separately in each field, and for each instrument. 
Double detections have been eliminated from the overlapping regions, removing detections with shorter exposure time from the source lists. 
Finally, source files obtained for each instrument have been cross-matched, creating therefore only one file, keeping only detections with S/N\,$>$\,2 in the total 0.5-10.0 keV range. 
The final catalogue has 1121 unique X--ray emitters, including sources detected in at least one of six energy bands. 

\begin{table*}[!ht]
\begin{center}
\caption{Selected energy bands, number of detected sources in each band, computed energy conversion factors, and flux properties of X-ray emitters in the SXDS field \label{tab_enrang_ecf_flux}}
\scriptsize{
\begin{tabular}{l c c c c c c}	
\noalign{\smallskip}
\hline
\hline
\noalign{\smallskip}
Band&Energy range&Num. of sources&MOS ecf&\textit{pn} ecf&Median flux&Limiting flux\\
&(keV)&&(10$^{11}$\,ct\,erg$^{-1}$\,cm$^2$)&(10$^{11}$\,ct\,erg$^{-1}$\,cm$^2$)&(10$^{-15}$\,erg\,cm$^{-2}$\,s$^{-1}$)&(10$^{-15}$\,erg\,cm$^{-2}$\,s$^{-1}$)\\
\noalign{\smallskip}
\hline
soft&0.5-2.0&1103&2.102&5.977&2.84&2.25\\
hard&2.0-4.5&1048&0.786&1.41&4.37&4.0\\
veryhard&4.5-10.0&976&0.193&0.433&5.59&5.0\\
total&0.5-10.0&1121&1.076&2.781&7.8&5.0\\
veryhard2&4.0-7.0&957&0.358&0.668&16.2&10.0\\
total2&0.5-7.0&1121&1.266&3.258&10.6&9.0\\
\noalign{\smallskip}
\hline
\end{tabular}
}
\end{center}
\end{table*}

\subsubsection{Comparison with Ueda et al. results}
\label{subsec_ueda}

\begin{figure}[ht!]
\centering
\begin{minipage}[c]{.51\textwidth}
\includegraphics[width=8.0cm,angle=0]{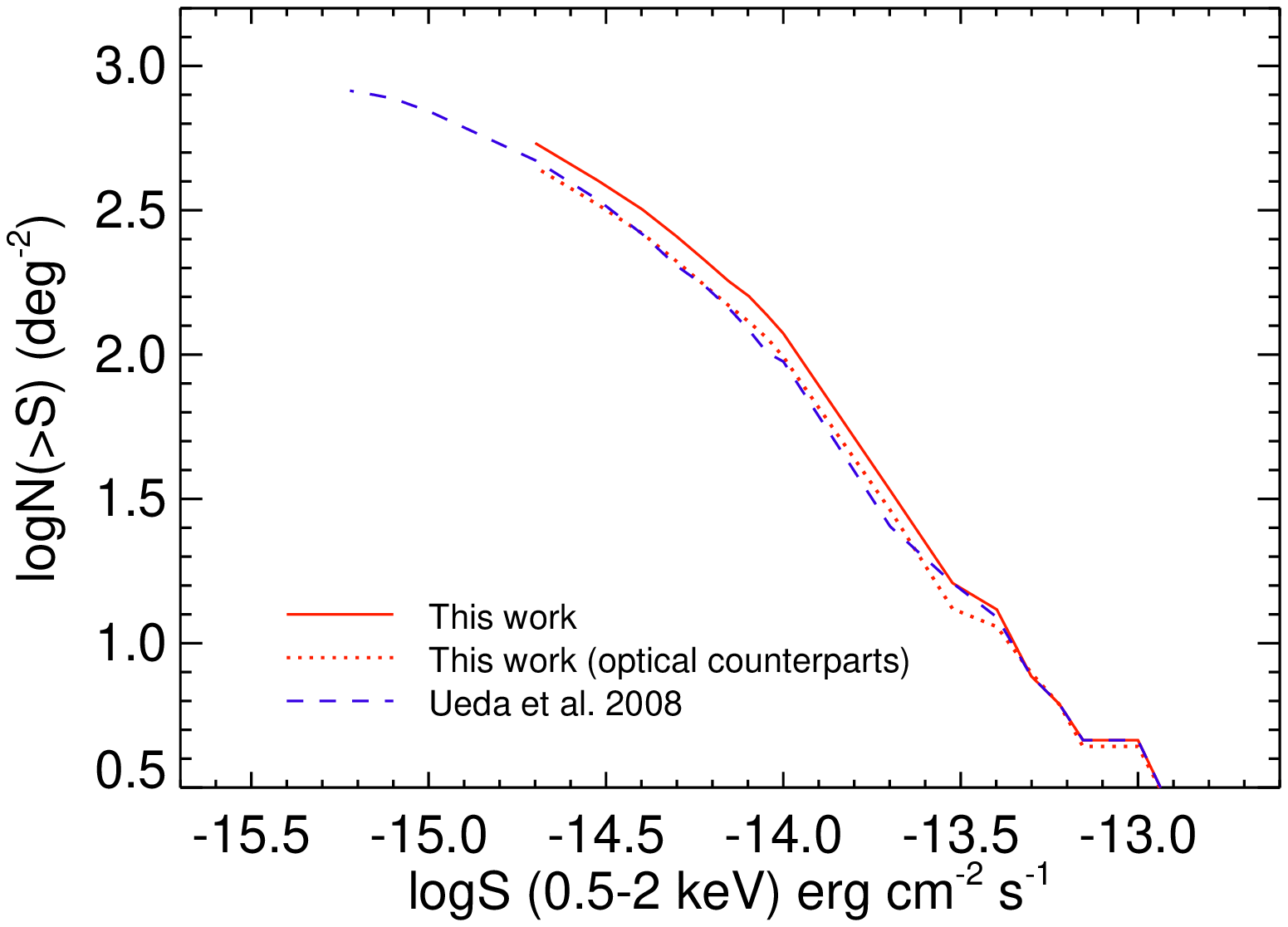}
\end{minipage}
\begin{minipage}[c]{.51\textwidth}
\includegraphics[width=8.0cm,angle=0]{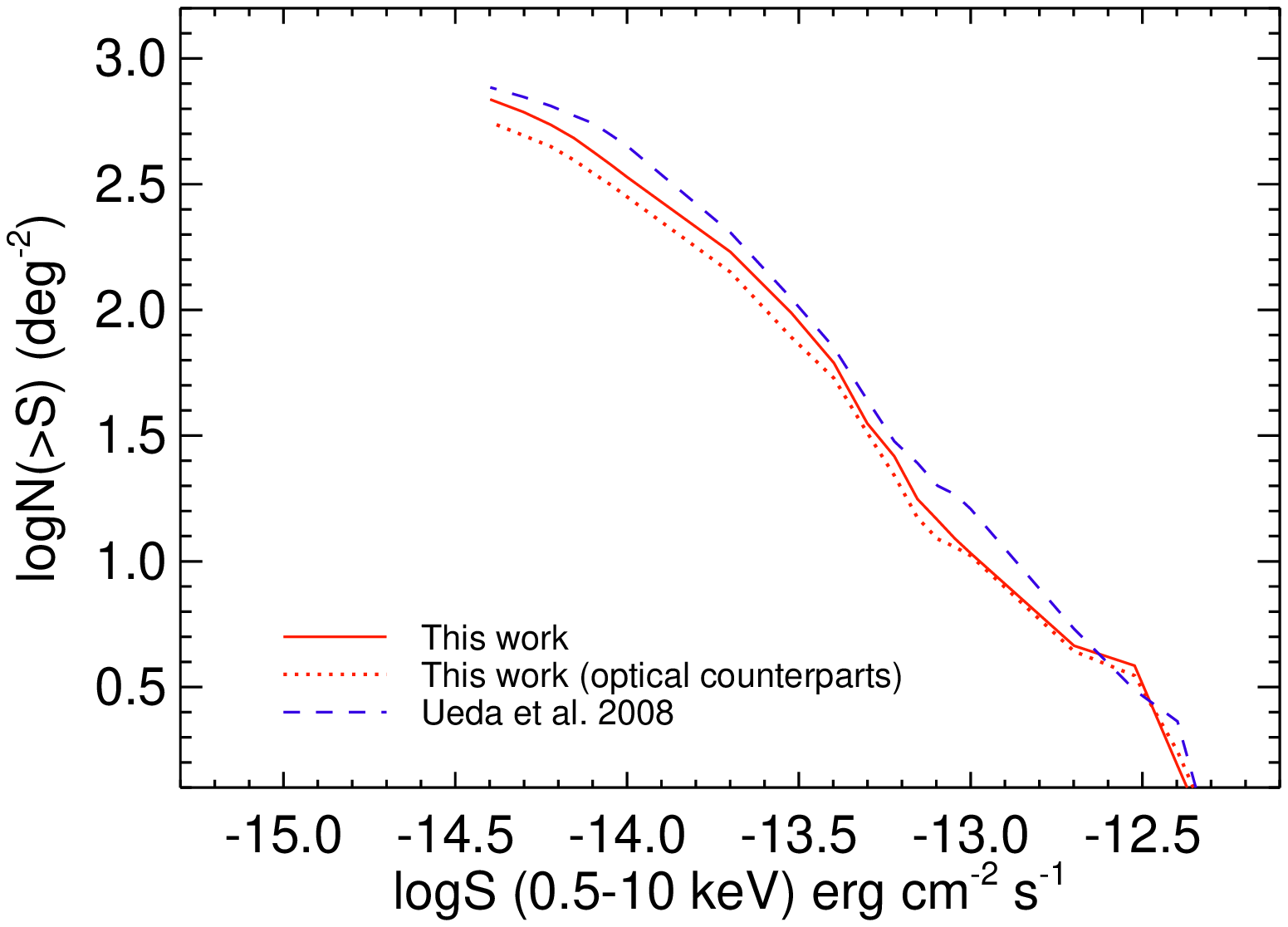}
\end{minipage}
\caption[ ]{Cumulative log\textit{N} - log\textit{S} functions for X--ray sources detected in the SXDS field in the soft 0.5\,-\,2\,keV (top) and total 0.5\,-\,10\,keV (bottom) bands. 
Solid and dotted red lines show the data obtained in this work, presenting the full X--ray sample and the sample with optical counterparts, respectively. The dashed blue line presents 
the data obtained by \cite{ueda08} in the same bands.}
\label{fig_lognlogs_ueda}
\end{figure}

\cite{ueda08} recently published a catalogue of X--ray emitters in the SXDS field. A total catalogue consists of 1245 X--ray sources detected in one of six selected energy bands: 0.3\,-\,0.5\,keV (ultrasoft),  0.5\,-\,2\,keV 
(soft), 2\,-\,4.5\,keV (medium), 4.5\,-\,10\,keV (ultrahard), 0.5\,-\,4.5\,keV (XID), or 2\,-\,10\,keV (hard). The data in the first four bands are publicly available. In general, the data reductions carried out in \cite{ueda08} 
and our works are quite similar. However, there are two main differences related with the source detection.
First, \cite{ueda08} performed the source detection to the summed \textit{pn} and MOS images, while we used individual images and finally summed detections from each camera, as explained above. 
Second, \cite{ueda08} adopted a detection likelihood threshold of 7 in a single band, while we used a more restrictive value, accepting only sources detected with a maximum likelihood above 14, and keeping only detections 
with S/N\,$>$\,2 in the total 0.5-10.0 keV range.\\
\indent Figure~\ref{fig_lognlogs_ueda} shows log\textit{N} - log\textit{S} relations in soft (0.5\,-\,2\,keV) and total (0.5\,-\,10\,keV) bands, where the total band for \cite{ueda08} data presents the sum of 
information obtained in their soft, medium and ultrahard bands. Although in the soft band the \cite{ueda08} data are more sensitive, our X--ray emitters with optical counterparts are in good correlation with the whole 
\cite{ueda08} sample. However, a higher density of sources has been detected in our full sample, probably due to the differences between source detection procedures explained above. On the other hand, 
limiting fluxes are in good correlation in the total band, and a higher density of sources has been detected in the \cite{ueda08} work, in comparison with both our full sample and optical counterparts, 
probably due to the less sensitive detection threshold used by the authors. \\
\indent Using a cross-matched radius of 5 arcsec, 857 counterparts (808 unique ones) have been found between \cite{ueda08} and our catalogues. Sources contained in the \cite{ueda08} catalogue that 
are missing in ours are in general fainter (e.g., around 90\% of these objects have fluxes below our limiting flux in the soft band). On the other hand, sources contained in our catalogue but not contained in the \cite{ueda08} 
catalogue are in general at lower S/N ratios (e.g., $\sim$\,50\% and $\sim$\,75\% of objects have S/N ratios between 2-3 and 2-4, respectively). Good correlation has been found for all counterparts between fluxes, with 
the linear correlation coefficients of 0.99, 0.97, 0.93, 0.96, and 0.99 in the 0.5\,-\,2\,keV, 2\,-\,4.5\,keV, 4.5\,-\,10\,keV, 0.5\,-\,10\,keV, and 0.5\,-\,4.5\,keV bands, respectively, where the correlation coefficient of 
1.0 presents a perfect linear correlation. In the soft band the \cite{ueda08} data are more sensitive, as mentioned above, while in the ultrahard and total bands limiting fluxes are in a good agreement. 
We are not able to provide the comparison between our and \cite{ueda08} X--ray luminosities, 
since they were not measured by the authors.\\

\subsection{Optical data}
\label{subsec_optdata}
Optical imaging observations were carried out with the Suprime-Cam installed at the Subaru Telescope
at Mauna Kea \citep{furusawa08}.
Five continuous sub-fields were observed, covering an area of 1.22\,deg$^2$ with a total integration time of 133 hours.
Each sub-field corresponds to a single Suprime-Cam field of view $\sim$\,34'\,$\times$\,27', in five broad-band filters B, V, $R_c$, $i'$,
and $z'$, with limiting AB magnitudes at 3$\sigma$ of: 28.4, 27.8, 27.7, 27.7, and 26.6, respectively. Publicly available catalogues have been downloaded
from the  SXDS web page\footnote{http://www.naoj.org/Science/SubaruProject/SXDS/index.html}, each of them containing 5-band aperture photometry (performed through
fixed 2 and 3\,arcsec apertures). The total number of objects are: 940,853, 1,002,561, 901,094, 899,484, and 842,590, in the B, V, $R_c$, $i'$,
and $z'$ bands, respectively. 

\subsection{Near-IR data}
\label{subsec_NIRdata}
We have used near-IR (NIR) data from the XMM-LSS survey (that encompasses the SXDS field), being part of the United Kingdom IR Telescope Deep Sky Survey (UKIDSS). 
Specifically, we have downloaded and used
the UKIDSS $JHK$ Second Data Release (DR2) catalogue\footnote{http://www.ukidss.org/archive/archive.html;
\citet{warren07}}. The photometric system is described in \cite{hewett06},
and the calibration is described in \cite{hodgkin09}. The AB limiting magnitudes at 2$\sigma$ are 23.4, 23.4, and 22.9 in $J$, $H$, and $K$ bands, respectively.

\subsection{Mid-IR data}
\label{subsec_MIRdata}
The SXDS field has been observed with the \textit{Spitzer} telescope as part of the XMM-LSS field, one of the areas surveyed within the \textit{Spitzer} Wide-area 
IR Extragalactic Legacy Survey (SWIRE).
The public IRAC catalogue has been downloaded from the SWIRE web
page\footnote{http://swire.ipac.caltech.edu/swire/astronomers/data\_access.html} and used in this work.
It provides a set of fluxes within five different aperture radii: 1.4, 1.9, 2.9, and 5.8\,arcsec for the four IRAC channels (3.6, 4.5, 5.8, and 8.0\,$\mu$m). Any source included in the catalogue, 
has been detected in both 3.6\,$\mu$m and 4.5\,$\mu$m bands. In order to convert the given
fluxes to magnitudes we have used the following equation:
\begin{equation} \label{eq_iracflux_to_mag}
{m[i] = 2.5 log_{10} (F_{zero}^{[i]}/F_{\nu}^{quot})},
\end{equation}
where $i$\,=\,3.6, 4.5, 5.8, and 8.0 are the four IRAC channels \citep{reach05}. $F_{\nu}^{quot}$ is the flux density of a source
from the calibrated images, and $F_{zero}^{[i]}$ are the zero-magnitude flux densities. The zero-magnitude
fluxes have been determined by \citet{reach05} integrating the Kurucz model spectrum of $\alpha$Lyr over the
IRAC pass-bands. The resulting zero-magnitude flux densities used in this work are 280.9\,$\pm$\,4.1,
179.7\,$\pm$\,2.6, 115.0\,$\pm$\,1.7, and 64.13\,$\pm$\,0.94 Jy in the 3.6, 4.5, 5.8, and 8.0 $\mu$m channels, respectively. The total SWIRE XMM-LSS catalogue has $\sim$\,250,700 objects. The AB limiting magnitudes at 
2$\sigma$ are 21.0, 21.1, 19.5, and 19.4 in 3.6, 4.5, 5.8, and 8.0\,$\mu$m bands, respectively.\\

\subsection{The catalogue of optical counterparts}
\label{subsec_Optcounterp}

Due to the very high number of detections in the optical bands (around 900,000 sources in both the $R_c$ and $i'$ bands, as noted in Section~\ref{sec_data}), there is a high probability of an erroneous 
identification when matching the X--ray and optical catalogues. In order to minimise such a risk, the optical catalogue
has been first filtered. Based on previous results in the GWS field \citep{povic09a}, the
optical flux has been constrained by assuming an upper limit of the X--ray-to-optical flux
ratio\footnote{Computed as $F_{\rm 0.5-4.5keV}/F_R$, where the optical flux $F_R$ has
been derived from the SExtractor auto magnitudes in the $R$ band; See Section~\ref{sec_colours} for more information and references}, X/O\,=\,18. We found this X/O ratio as a reasonable limit, since 
no sources have been found above this value in the GWS field. With this cut, and with the minimum X--ray flux of 8.9\,$\times$\,$10^{-16}$\,erg\,cm$^{-2}$s$^{-1}$ detected in the SXDS field, we obtained an 
upper limit in the optical range, and excluded all optical detections having AB magnitude $R_c$\,$>$\,27.17. After applying this filter, the resulting catalogue is reduced from 901,094 $R$ band sources to 614,857. 
This approach has proved to be consistent with the SXDS sample, since none of the X--ray sources with 
F$_{\rm 0.5-4.5keV}$\,$\le$\,6.0\,$\times$\,$10^{-15}$\,erg\,cm$^{-2}$s$^{-1}$ show X/O\,$>$\,18,
while only about 5\% of the total number of sources exceed this limiting X/O value. \\
\indent After filtering the optical catalogue, we followed the same cross-matching procedure as explained in \cite{povic09a}. Comparing the number of unique counterparts and the number of multiple matches, 
the radii between 2$''$ and 3$''$ have been found to be the best compromise, selecting always as an optical counterpart a source closest to the X--ray detection. 
After applying the statistical methodology from de Ruiter \citep{deruiter77} a radius of 3$''$ has been selected, obtaining a completeness of 99.9\% and a reliability of 76.2\%. The total number of optical counterparts 
obtained with this radius is 808. Additionally, for 
objects having multiple matches inside the selected radius, we performed the \cite{sutherland92} methodology measuring the reliability for all possible matches, and finally selecting as an optical counterpart an object with 
the highest probability. After this the number of possible fake identifications drops to $\sim$\,6\%, where multiple matches are detected with similar probabilities.
Regarding the $\sim$\,300 X--ray sources not optically matched, 80 objects reside in the area not covered by optical observations; the rest of the objects might have been lost through optical catalogue filtering, 
they may be optically 
obscured sources, or some have resulted from fake X--ray detections (see Section~\ref{subsec_Xdata}).

\subsection{Photometric redshifts and k-corrections}
\label{subsec_zphot_kcorr}

\begin{figure}[ht!]
\centering
\includegraphics[width=8.0cm,angle=0]{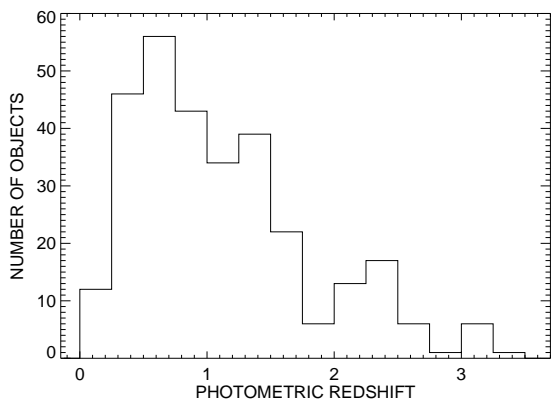}
\centering
 \protect\caption[ ]{Distribution of final photometric redshifts of X--ray emitters with optical
counterparts. 
\label{fig_zphot_sxds}}
\end{figure}

We used the Bayesian code ZEBRA \citep{Feldmann} to compute photometric redshifts of the optical
counterparts. Due to the small number of available spectroscopic redshifts used in the analysis (for only 15 sources obtained from \cite{aretxaga07}, and references therein), as 
an additional check, photometric redshifts were also computed using the HiperZ code \citep{bolzonella2000}. In both codes the templates from the 
SWIRE\footnote{http://www.iasf-milano.inaf.it/$\sim$polletta/templates/swire\_templates.html} template library were implemented and used, since they include the templates of 
AGN/QSO sources \citep{polletta07}. Comparing the results from ZEBRA and HyperZ, only those objects for which the redshift results from both codes agree within less 
than 0.1 for z\,$<$\,1 and 0.2 for z\,$\geq$\,1 were accepted. Finally, we derived high-quality photometric redshift information for 308 objects shown in 
Figure~\ref{fig_zphot_sxds}. The mean value of the final photometric redshifts is 1.17\,$\pm$\,0.15. Of those 86\% of objects have photometric redshift errors below 20\%.

In order to obtain k-corrections of the optical fluxes we have used the \textsc{idl} routine
\textsc{kcorrect} \citep{blanton}. The X--ray fluxes, on the other hand, were k-corrected by assuming a standard power law SED with $\Gamma$\,=\,1.8. A rough classification between 
X--ray unobscured and obscured AGN has been made using the hardness ratio
HR$(2-4.5keV/0.5-2keV)$ (see Section~\ref{subsec_colour_nucty}), which is quite sensitive to absorption \citep{della04,caccianiga04,dwelly05}; those objects with HR$(2-4.5keV/0.5-2keV)$\,$\le$\,0.35
(some 53\%) have been considered as unobscured, and no intrinsic absorption has been applied to the power law
SED; otherwise the objects have been considered as obscured and a fixed intrinsic absorption 
\nh\,=\,1.0\,$\times$\,10$^{22}$\,cm$^{-2}$ has been included \citep[e.g.,][]{silverman05,younes11}.

\textbf{\subsection{Selected sample}}
\label{subsec_sample_selection}

\begin{figure}[ht!]
\centering
\includegraphics[width=9.0cm,angle=0]{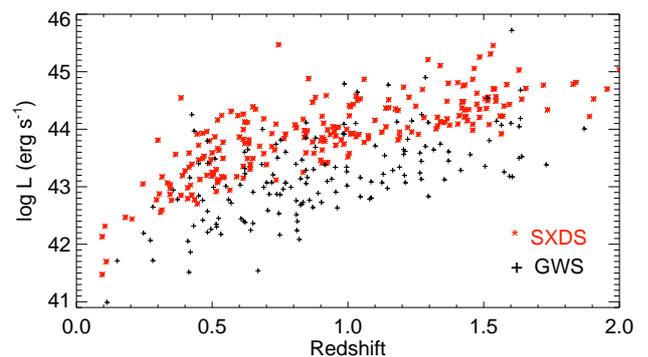}
\centering
\protect\caption[ ]{Relation between the X--ray luminosity in the 0.5\,-\,7.0\,keV band and redshift for a sample of AGN in the SXDS (red stars) and GWS (black crosses) fields.
\label{fig_Lz_plane}} 
\end{figure}

After performing \cite{deruiter77} and \cite{sutherland92} methodologies when cross-matching X--ray and optical catalogues, and after considering only those objects with high-quality photometric 
redshift information, we remained with 308 sources in our sample. Of these, 262 sources (85\%) have photometric redshifts $\le$\,2.0. We obtained the morphological classification for all 308 objects 
(see Section~\ref{sec_morph}), but we carried out all further analysis for a sample of objects with redshifts z\,$\le$\,2.0 and with redshift errors below 20\%. For this sample we derived X--ray luminosities 
in all selected energy ranges. The average luminosity of logL$_X$\,=\,43.7\,erg\,s$^{-1}$ has been measured in the 0.5\,-\,7.0\,keV band. Figure~\ref{fig_Lz_plane} shows the luminosity\,-\,redshift plane for 
X--ray detected AGN in the SXDS and GWS fields at z\,$\le$\,2.0. \\

\begin{figure}[ht!]
\centering
\begin{minipage}[c]{.51\textwidth}
\includegraphics[width=8.0cm,angle=0]{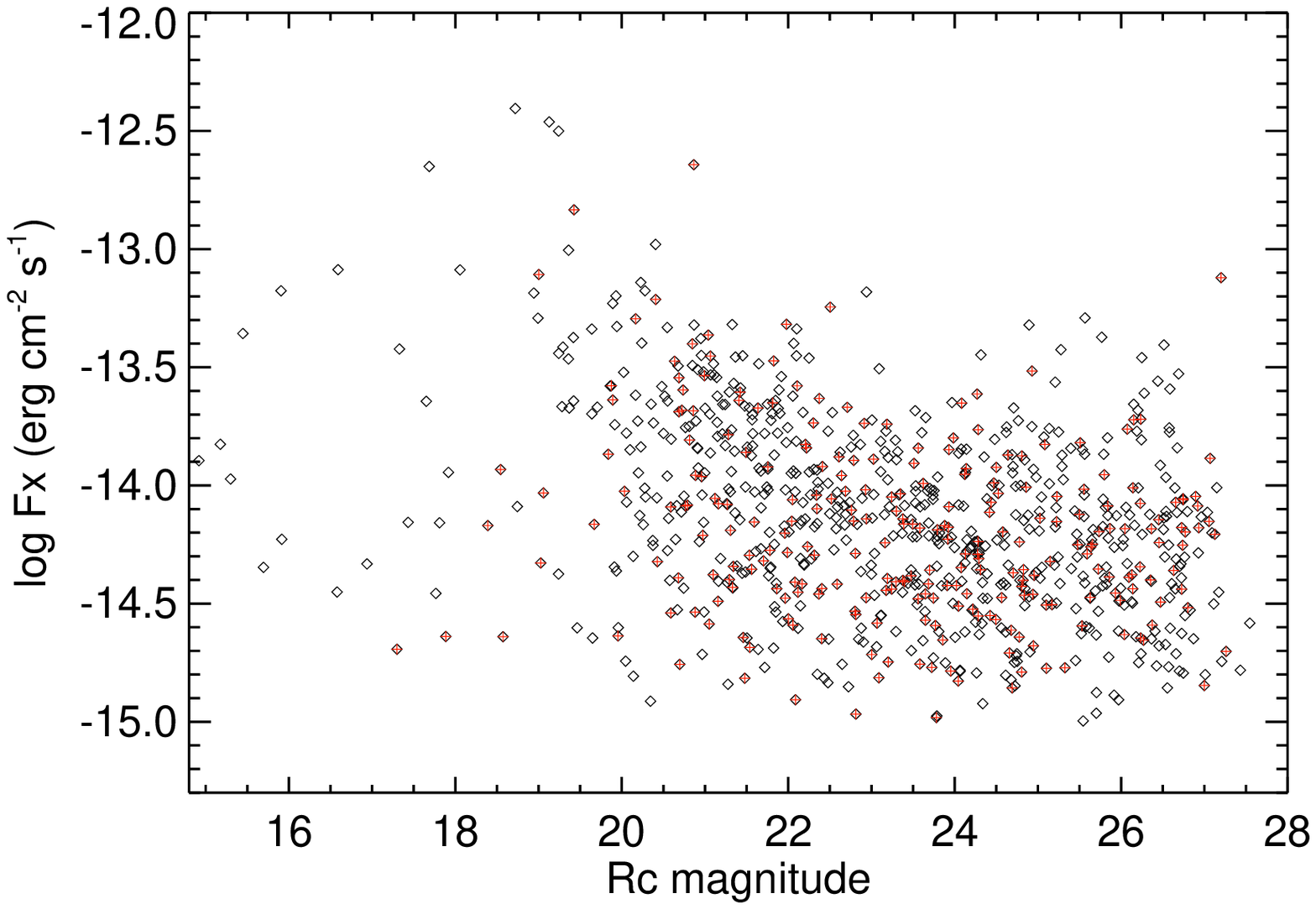}
\end{minipage}
\begin{minipage}[c]{.51\textwidth}
\includegraphics[width=8.0cm,angle=0]{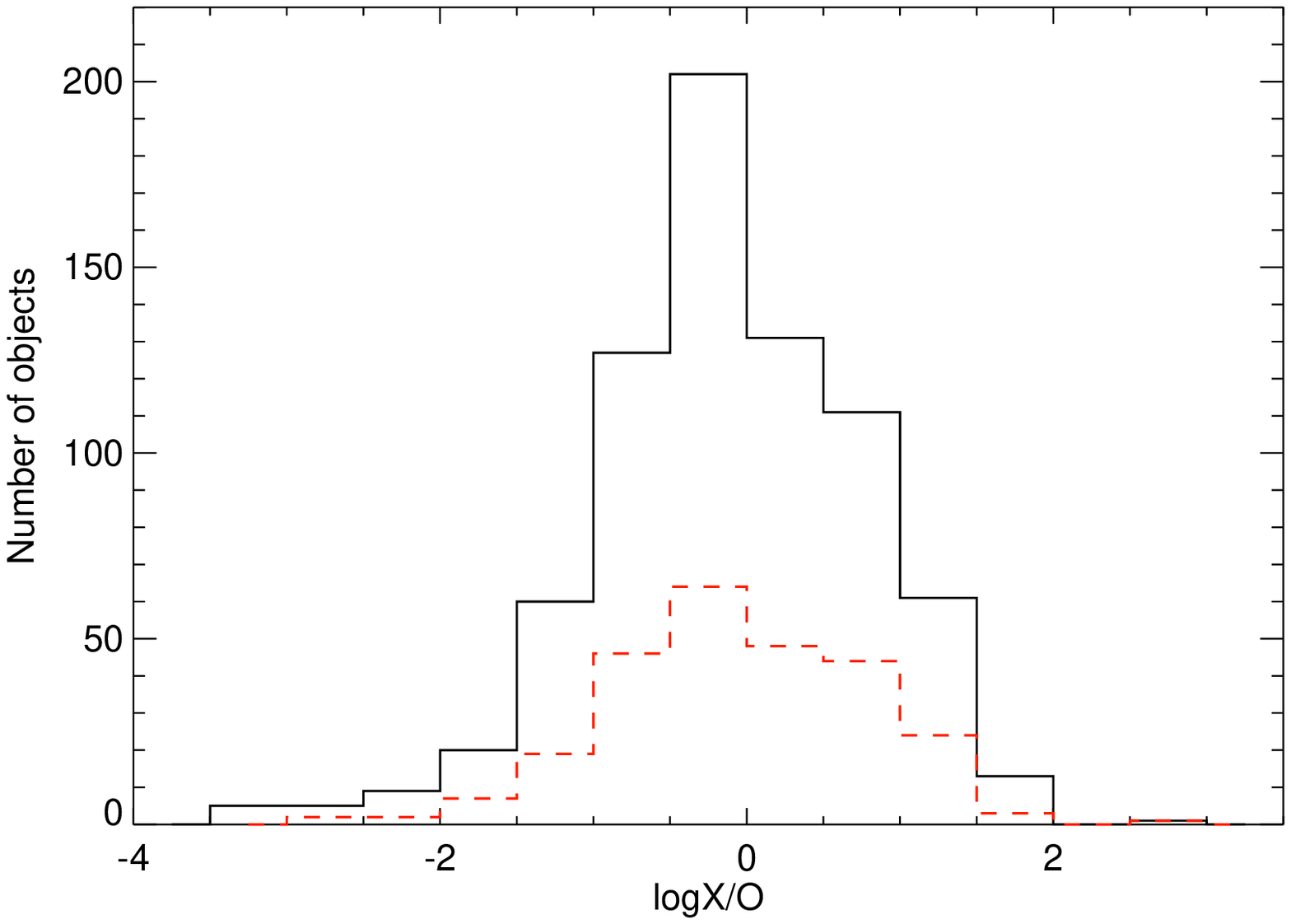}
\end{minipage}
\caption[ ]{\textit{Top:} Relation between the X--ray flux in the 0.5\,-\,4.5\,keV energy range and the AB R$_c$ magnitude of the full matched sample (black diamonds) and of the selected sample (red crosses). 
\textit{Bottom:} Distributions of X--ray-to-optical flux ratio of the full matched sample (black solid line) and of the selected sample (dashed red line).
\label{fig_XO_Fxrmag_comparisons}}
\end{figure}

The selected z\,$\le$\,2.0 sample presents around 25\% of the initial X--ray sample, and 32\% of the optically matched sample. In order to see if this sample is representative of the full X--ray population or of the 
matched one, we compared their X--ray flux distributions in all energy bands. Figure~\ref{fig_fx_distribution} shows these distributions in five energy bands for the parent X--ray sample and the analysed one. The selected 
sample seems to follow the distribution of full X--ray population in all observed bands quite well. Performing Kolmogorov-Smirnov statistics we obtained the probability parameter of 0.7 and 
0.97\footnote{Probability can take values from 0.0 to 1.0, where 1.0 means that the compared distributions are completely identical.} in the 0.5\,-\,4.5\,keV band 
(energy range used in AGN selection; see Section~\ref{sec_colours}) when comparing the analysed sample with the full X--ray population and with the full optically matched sample, 
respectively. Moreover, we compared X--ray and optical properties of the selected and full optically matched samples. Figure~\ref{fig_XO_Fxrmag_comparisons} (top) shows the relation between 
the X--ray flux in the 0.5\,-\,4.5\,keV energy range and the AB R$_c$ magnitude for these two samples. As again can be seen, the selected z\,$\le$\,2.0 sample covers the entire range of X--ray fluxes 
in comparison with the matched one. For a selected sample, there seems to be a shallow bias towards fainter sources in the optical range. However, when comparing the X--ray-to-optical flux ratios 
(Figure~\ref{fig_XO_Fxrmag_comparisons}, down panel), the selected sample represents the full matched one quite well. For these two samples, the average logX/O ratios are 0.05 and -0.07, respectively.

To summarise, the z\,$\le$\,2.0 selected sample is a small fraction ($\sim$\,25\%) of the full X--ray population (as mentioned above), but seems to follow the distributions of the initial X--ray and optically matched 
samples quite well. \\

Finally, in order to improve statistics, we included our data from the Groth-Westphal Strip (GWS) field when analysing the colour-magnitude relations. 174 objects were used, having photometric redshifts $\le$\,2.0 and 
redshift errors below 20\%. These data are deeper in X--rays in comparison with SXDS data, having X--ray limiting flux of 0.48\,$\times$\,10$^{-15}$\,erg\,cm$^{-2}$\,s$^{-1}$ in the 0.5\,-\,7.0\,keV energy band. The detailed 
description of data, morphological and nuclear classifications can be found in \citet{povic09a}. A small comparison between the luminosities in the SXDS and GWS fields is also presented below (see Figure~\ref{fig_Lz_plane} and 
Section~\ref{subsec_sample_selection}).
 
\section{Morphological classification and analysis}
\label{sec_morph}
\subsection{Methodology}
\label{sec_morph_method}

The morphology of the optical counterparts of X--ray emitters in the SXDS field was tested using SExtractor \citep{bertin96} and galSVM \citep{huertas08,huertas09} 
codes. Moreover, as an additional check, a visual classification was performed as well.

First, compact objects were separated from extended ones. The SExtractor CLASS\_STAR parameter was used to this end. This parameter permits the rough classification of objects as 
compact (QSOs, AGN with the weak host galaxies, stars, etc.) or extended. The computed value depends on the seeing, the peak intensity of the source and its isophotal area. The parameter 
ranges between 0 (for an extended object like a galaxy) and 1 (for a point-like object). A source has been deemed as compact if its CLASS\_STAR parameter is $\ge$\,0.9.

In order to separate between early- and late-type X--ray emitters (see Section~\ref{sec_morph_class} below), galSVM code has been used \citep{huertas08,huertas09}. galSVM\footnote{http://gepicom04.obspm.fr/galSVM/Home.html}
is a freely available code written as an IDL library that, combined with
the also freely available library libSVM\footnote{Support Vector Machines
library (libSVM), written by Chih-Chung Chang \& Chih-Jen Lin, is an integrated program
for support vector classifications, regressions, and distribution estimations;
it supports the multi-parameter classifications.} enables a morphological classification of galaxies in an automated
way using support vector machines (SVM\footnote{Support Vector Machines (SVM)
are a group of supervised learning methods that can be applied to classification or regression.
Support Vector Machines represent an extension to non-linear models of the generalised portrait algorithm developed by Vladimir Vapnik (1995).}). 
This code represents a great improvement within non-parametric methods for morphological classification, being especially useful when dealing with low-resolution, and high-redshift data. 
It makes it possible to use a sample of objects with known morphology, to move these objects to the redshift and image quality of the sources that have to be classified, to 
test simultaneously different morphological parameters in order to find the best set for the morphological classification, and to use non-linear boundaries in order to determine the 
separation regions between different types. Different morphological parameters have been suggested and used in morphological classification over recent years
\citep{abraham96,kent85,bershady00,abraham03,conselice03,lotz04}, each of them providing different types of information about the structure of a galaxy. Before galSVM it 
was not possible to use more than 3 parameters simultaneously, forcing boundaries between different Hubble types to be linear (2D lines or hyper-planes). Therefore, besides being 
useful when dealing with high redshift and low resolution data, there are two main advantages of galSVM compared to other codes: first, it can use a list of morphological parameters 
simultaneously, and second, it can find and use non-linear boundaries in order to perform the final morphological classification. It has been shown that using galSVM to classify galaxies into 
two main morphological types (early- and late-type galaxies), provides a more reliable classification when compared to other non-parametric methods \citep{huertas08}. Therefore, since we are dealing 
with low-resolution and high-redshift data, we decided to use the galSVM code in order to separate between early- and late-type X--ray emitters in the SXDS field. galSVM requires the availability 
of redshift information for all examined objects. Therefore, morphology has been tested on the 308 X--ray sources with known photometric or spectroscopic redshifts. To perform the analysis, the $i'$ 
optical band was chosen due to its higher S/N ratio.

We followed the complete and standard galSVM procedure, as described in \cite{huertas08}, in order to separate between early- and late-type X--ray emitters with optical counterparts:
\begin{itemize}
\item[1.] We built a set of simulated galaxies, using a catalogue of local galaxies with known Hubble type (obtained by visual classification), redshift, total flux, and half luminosity radius. 
The \citet{tasca11} catalogue was used for this purpose, containing 1504 visually classified local galaxies from the SDSS survey, observed in the $g$ photometric band, which corresponds 
to the rest-frame\footnote{As noted by \cite{huertas08}, selecting as the rest-frame band the training sample has three main advantages: \textit{(i)} it is less affected by k-correction 
effects, \textit{(ii)} it does not introduce any modelling effect, since the used galaxies are real, and \textit{(iii)} it is possible to work with seeing limited data (as the training set is 
built to reproduce the observing conditions and physical properties of the sample to be analysed, but is classified using well-resolved images).} band of the SXDS $i'$ sample used with the average 
redshift of 0.8. Besides the local catalogue, subimages of all local galaxies and PSFs, necessary to run galSVM, have been gathered from \citet{tasca11} as well. 
For each of the five SXDS fields observed in the optical $i'$ band, 
we provided the magnitude and redshift distributions of our X--ray emitters with optical counterparts. Moreover, we run SExtractor on our real images 
in order to obtain the catalogue of all possible detections, as well as mask images. For every galaxy stamp used for training the morphology, galSVM generates a random pair of magnitude and redshift values with a 
probability distribution that matches that of our real data, in order to place the training galaxy in a real, high-redshift background image of our sources. Using the real images and SExtractor 
catalogues, the locations devoid of objects where the trained sample of galaxies could be placed are searched randomly, and then a SExtractor mask image is used to eliminate close objects. 
\item[2.] After placing the sample of local galaxies in the high-redshift background of our real sample, we measured a set of morphological parameters of the training sample. 
\item[3.] We trained a SVM with a fraction of 
600 local galaxies and used the other fraction
of 904 local galaxies to test the accuracy\footnote{Accuracy represents the success rate in the morphological classification of the training sample; see \cite{huertas08} for more information.}, 
and to estimate errors. Then we repeated steps two and three to test different sets of morphological 
parameters and to choose the one that gives the most reliable morphological classification, which will be used afterwards for the classification of our real sample. As already mentioned, the galSVM 
accuracy parameter has been used as an indicator of classification reliability, but aside from the accuracy, the distributions of the obtained morphological parameters have been tested as well. 
It has been seen that there is a trend for bright objects to be identified as late-types with very high probabilities, increasing the value of the mean 
probability. Therefore, all brightness parameters (as mean surface brightness, magnitude, flux) have been excluded from the initial set of parameters 
involved in the morphological classification. Table~\ref{tab_morph_class_set_sxds} shows the final set of parameters used to separate between early- and late-type galaxies. For this set of 
parameters an accuracy of 70\% is obtained (see Section~\ref{sec_morph_class} below). \\
\item[4.] A set of morphological parameters has been obtained for all X--ray emitters with optical counterparts, and using the parameters from 
Table~\ref{tab_morph_class_set_sxds}, morphological classification using galSVM was derived, correcting it for possible systematic errors detected in the testing steps.\\
\end{itemize}

\begin{table*}[!ht]
\begin{center}
\caption{Set of parameters used by galSVM for the final morphological
classification of X--ray emitters in the SXDS field
\label{tab_morph_class_set_sxds}}
\scriptsize{
\begin{tabular}{l l}
\noalign{\smallskip}
\hline\hline
\noalign{\smallskip}
ELONGATION$^1$&CLASS\_STAR\\
ABRAHAM CONCENTRATION INDEX$^2$ ($\alpha$\,=\,0.3)&ASYMMETRY$^2$\\
SMOOTHNESS$^3$&GINI COEFFICIENT$^4$\\
M$_{20}$ MOMENT OF LIGHT$^5$&BERSHADY-CONSELICE CONCENTRATION INDEX$^6$\\
\noalign{\smallskip}
\hline
\hline
\end{tabular}
}
\normalsize
\rm
\end{center}
\begin{flushleft}
\scriptsize{
$^1$ Ratio between the major and minor axes \citep[A\_IMAGE and B\_IMAGE SExtractor parameters;][]{bertin96}\\
$^2$ \cite{abraham96}\\
$^3$ \cite{conselice03}\\
$^4$ \cite{abraham03}\\
$^5$ \cite{lotz04}\\
$^6$ \cite{bershady00}
}
\end{flushleft}
\end{table*}

\subsection{Morphology of X--ray emitters}
\label{sec_morph_class}

The morphological classification obtained, gives the result that early-type galaxies include elliptical and lenticular (although in some cases early-type spirals between S0 and Sa galaxies might be classified as early-type as 
well), 
while spiral and irregular galaxies have been classified as late-type. The output classification assigns to each galaxy a class label
and a probability of belonging to a given class.
When dealing with a 2-class problem \citep[early- and late-type
classification; see ][]{huertas08}, the probability $p$ will be:
$p_{early-type}$\,=\,1\,-\,$p_{late-type}$.
Therefore, two probabilities are measured:
$p_1$\,=\,$p_{early-type}$ and
$p_2$\,=\,$p_{late-type}$, where for galaxy to be considered as early- or late-type the probability $p_1$ or $p_2$ must be higher than 0.5, respectively. The probability parameter is used to assess
the accuracy of the morphological classification.
There is a clear correlation between the probability threshold and the
number of correct identifications: the accuracy clearly increases when
the considered probability is higher. In \citet{huertas09}
it has been shown that selecting objects with a probability between 0.5 and 0.6
yields a mean accuracy of around 58\%, while objects with probabilities
greater than 0.8 are classified with nearly 90\% accuracy.

\begin{figure*}[ht!]
\centering
\includegraphics[width=16cm,angle=0]{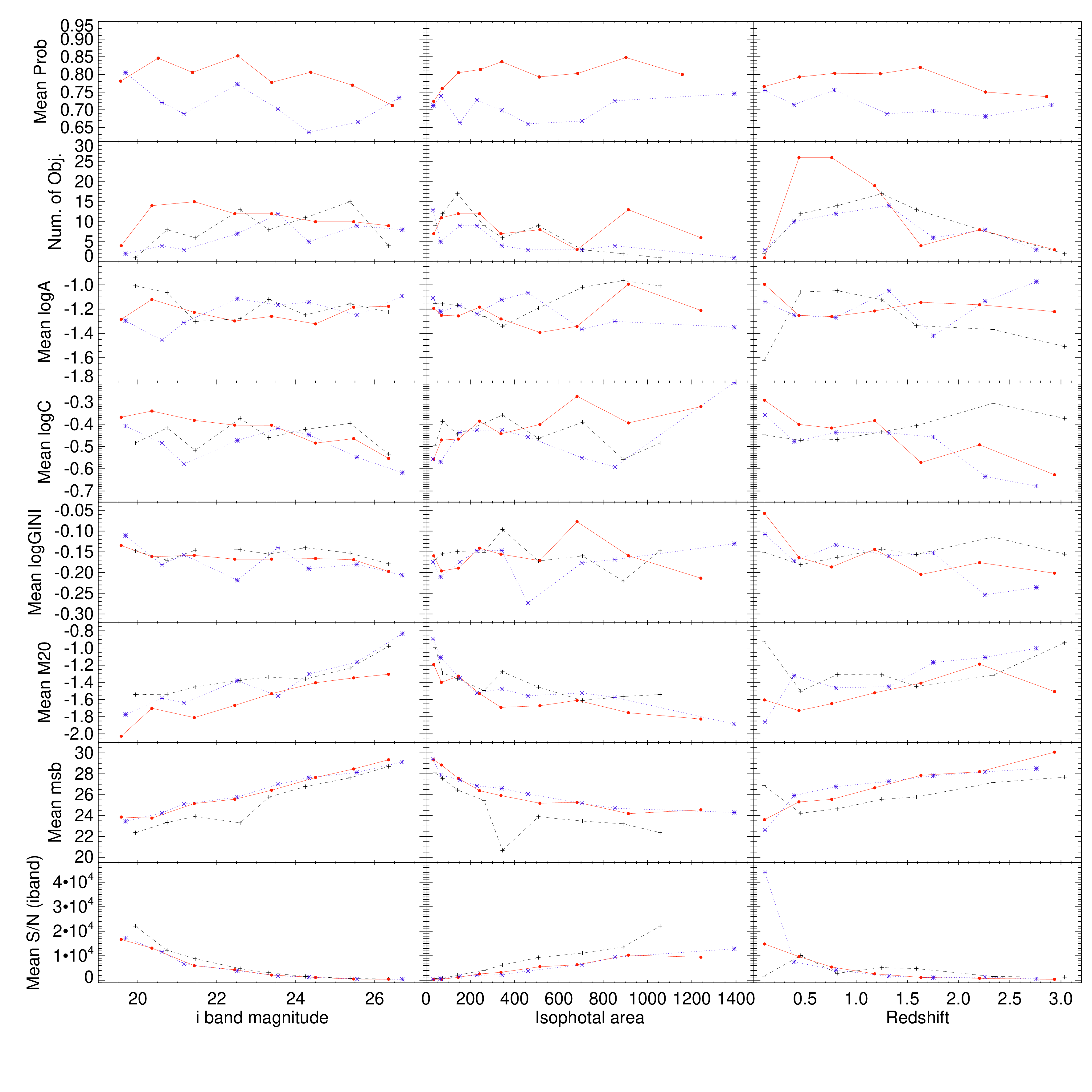}
\protect\caption[ ]{\textit{From top to bottom:} Relationship between mean values of galSVM probabilities, number of objects, asymmetry index, Abraham concentration index, Gini coefficient, M$_{20}$ moment of light, 
mean surface brightness, S/N in i band, and (\textit{from left to right:}) apparent magnitude in $i'$ band, isophotal area, and redshift. Compact objects are marked with black crosses and dotted lines, early-type galaxies 
with red spots and solid lines, and finally, late-type galaxies with blue stars and dashed lines.
\label{im_morph_vs_zmagiso_sxds}}
\end{figure*}

As already mentioned in Section~\ref{sec_morph_method}, the SExtractor STAR\_CLASS parameter has been used to separate between compact and extended (host-dominated) objects. 22\% of the total
number of objects have been detected as compact, having a STAR\_CLASS
parameter $\ge$\,0.9. Of those objects detected as early-type (with STAR\_CLASS\,$<$\,0.9 and $p_1$\,$>$\,0.5; 53\% in total), 56\% have probabilities $\ge$\,0.75. Conversely, of objects identified as late-type 
(with STAR\_CLASS\,$<$\,0.9 and $p_2$\,$>$\,0.5; 18\% in total) only 30\% are found
with probabilities $p_2$\,$\ge$\,0.75. A population of unclassified objects is 7\% in total.\\
\indent Several trends of the mean probability values have been tested
(for both early- and late-types) with redshift, magnitude, and isophotal area
(see Figure~\ref{im_morph_vs_zmagiso_sxds}, upper panels). Since the trends of morphological parameters with distance, brightness, and size are also present in the training sample, the algorithm should be aware of such 
trends and should be able to take them into account in the final classification. 
For $p_1$ probabilities (early-type objects)
there is only a weak trend with these parameters. Yet for
$p_2$ probabilities (late-type objects) the trends seem to be more significant.
This means that low $p_2$ probability objects (e.g., $p_2$\,$<$\,0.7)
still have a big chance of being one of the late-types, but are detected with low
probabilities since they are more distant, smaller, and/or fainter sources.

\begin{figure*}[ht!]
\centering
\begin{minipage}[c]{.49\textwidth}
\includegraphics[width=8cm,angle=0]{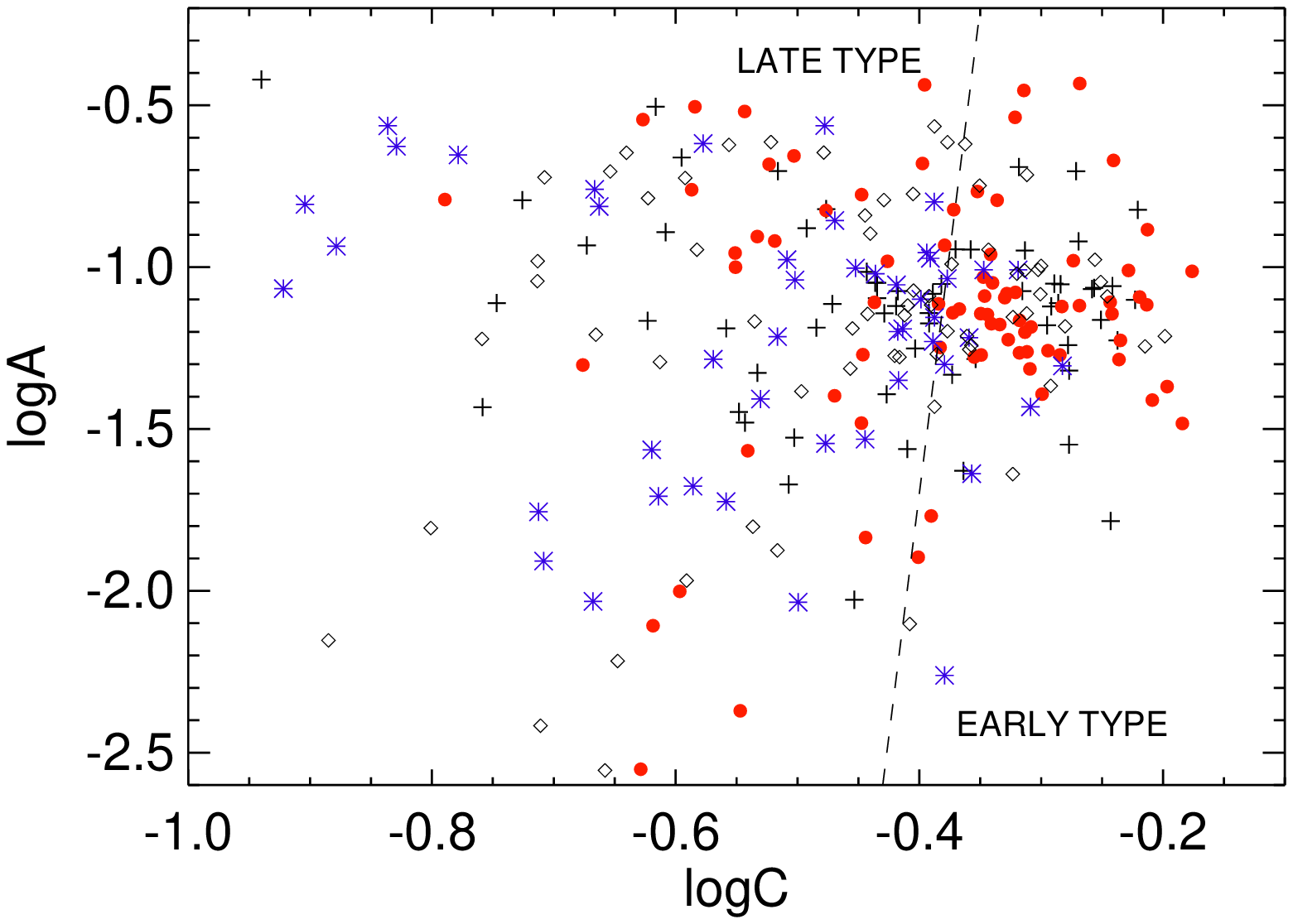}
\centering
\protect{(a)}
\end{minipage}
\begin{minipage}[c]{.49\textwidth}
\includegraphics[width=8cm,angle=0]{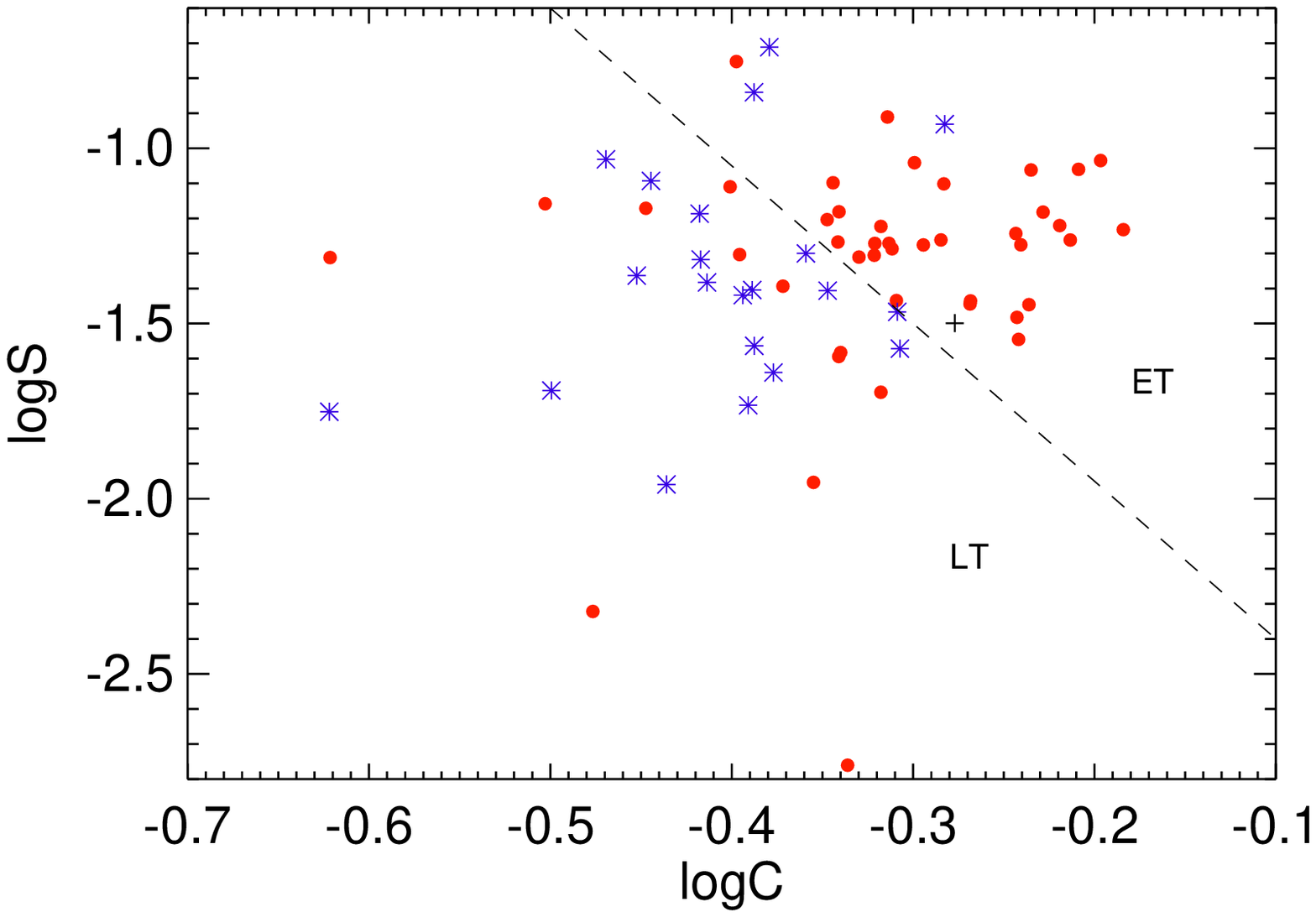}
\centering
\protect{(b)}
\end{minipage}
\begin{minipage}[c]{.49\textwidth}
\includegraphics[width=8cm,angle=0]{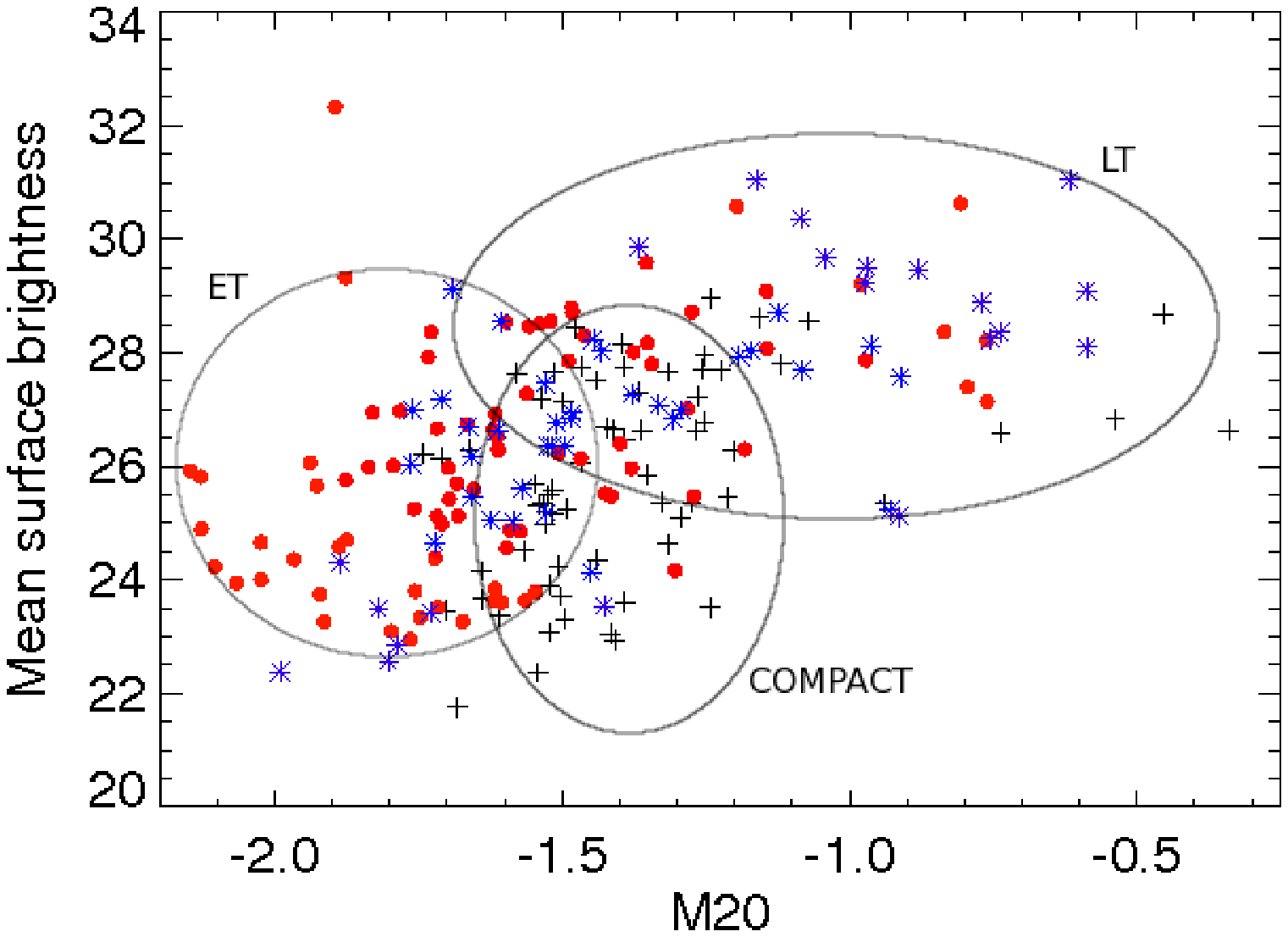}
\centering
\protect{(c)}
\end{minipage}
\begin{minipage}[c]{.49\textwidth}
\centering
\includegraphics[width=8cm,angle=0]{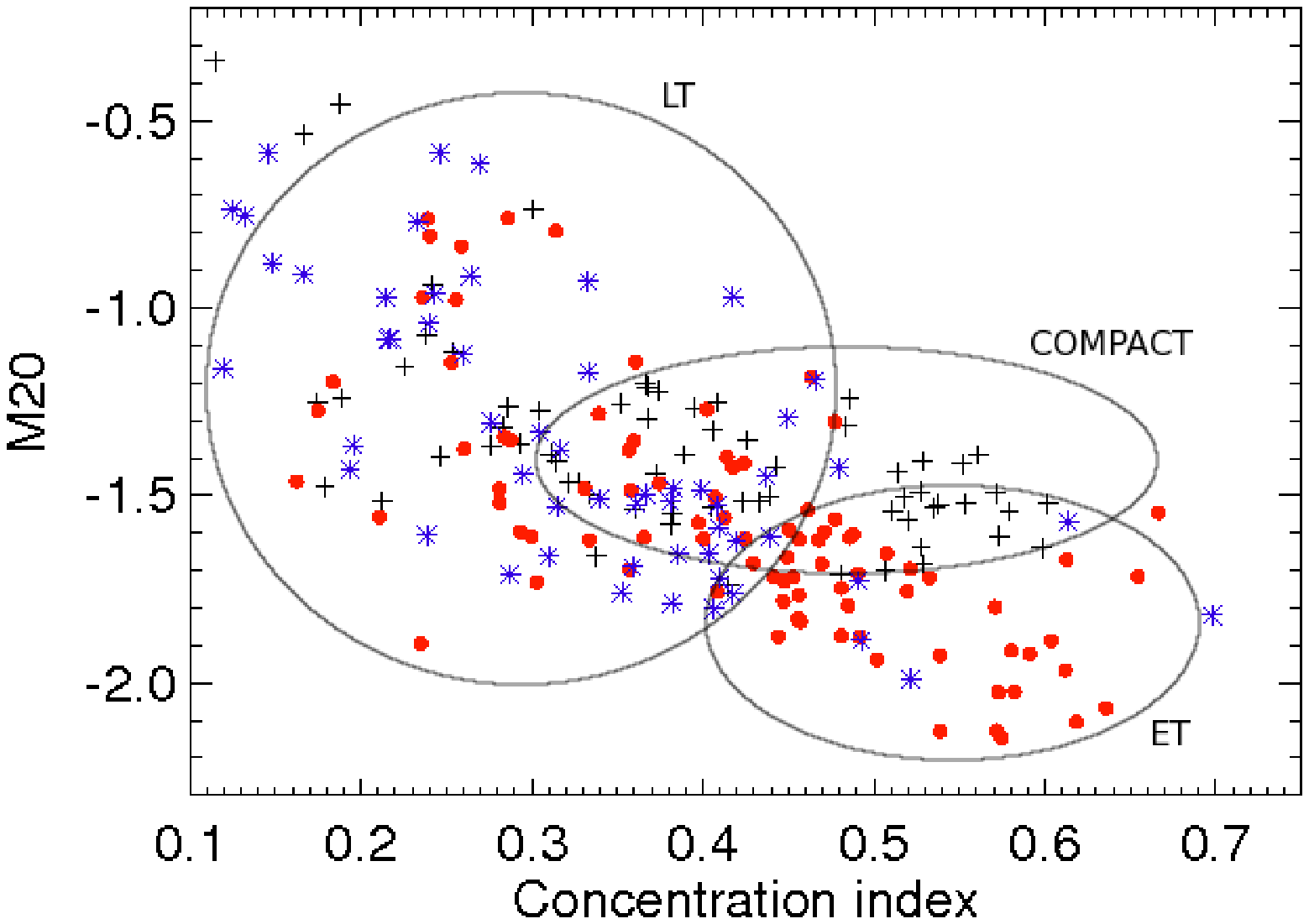}
\centering
\protect{(d)}
\end{minipage}
\begin{minipage}[c]{.49\textwidth}
\includegraphics[width=8cm,angle=0]{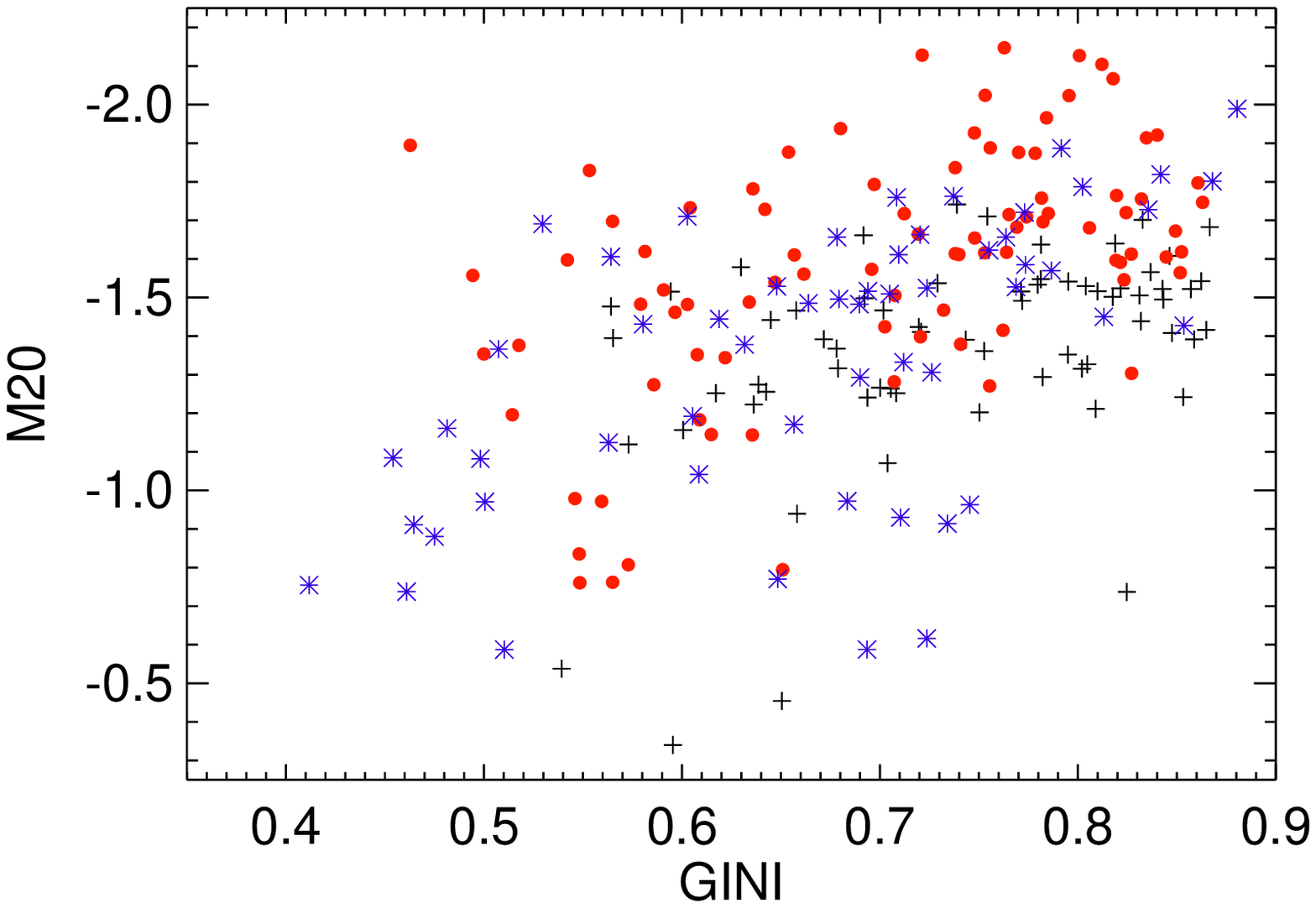}
\centering
\protect{(e)}
\end{minipage}
\begin{minipage}[c]{.49\textwidth}
\centering
\includegraphics[width=8cm,angle=0]{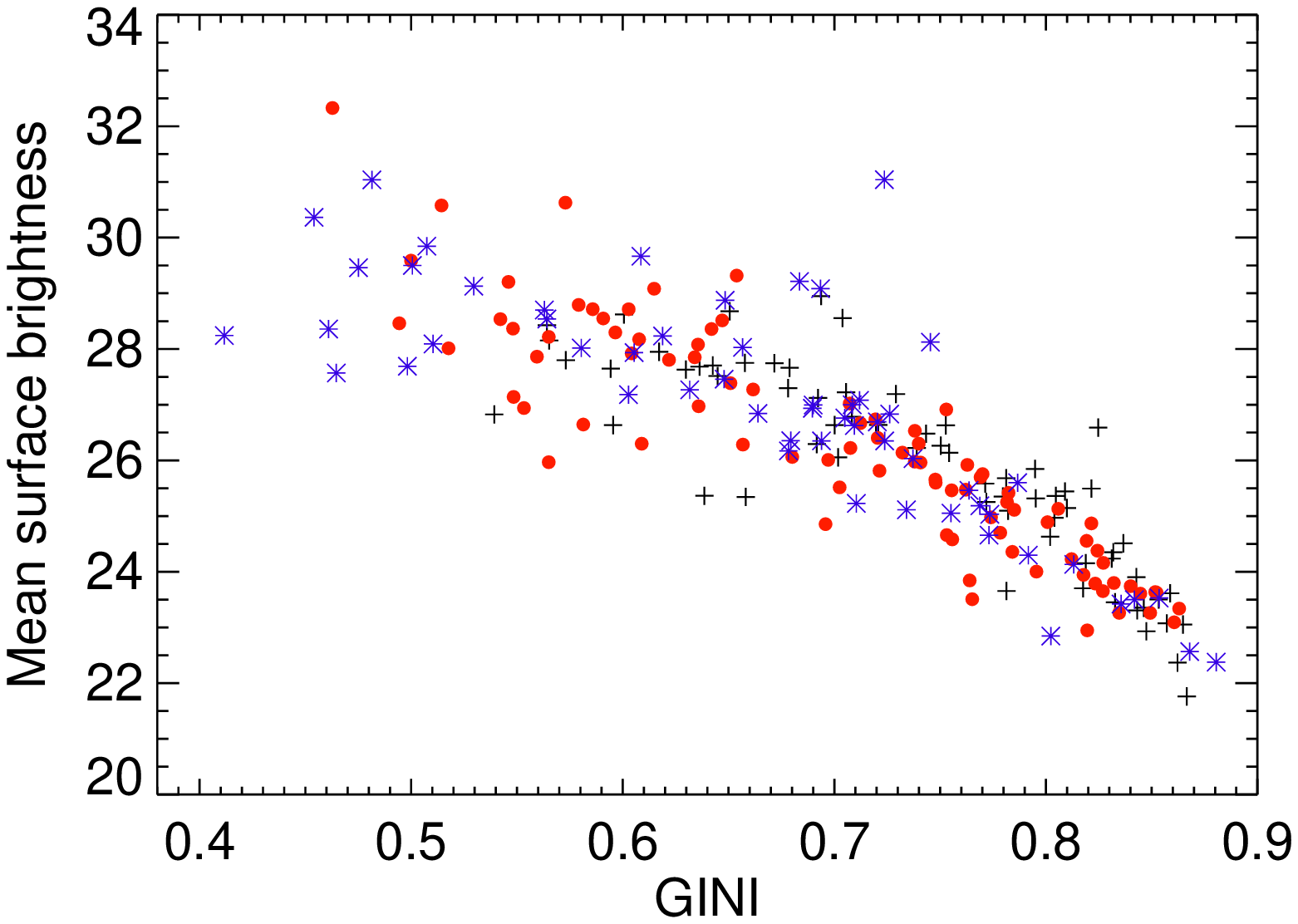}
\centering
\protect{(f)}
\end{minipage}
\protect\caption[ ]{\textit{Top left:} Relation between asymmetry and Abraham concentration index. Red dots represent the early-type galaxies (E/S0, S0/Sa) classified with probabilities
$p_1$\,$\ge$\,0.75, while the blue stars represent all late-type galaxies having $p_2$ probabilities above 0.5. Black crosses represent compact objects (all sources with
CLASS\_STAR\,$\ge$\,0.9). Open diamonds are sources with 0.5\,$<$\,$p_1$\,$<$\,0.75, which might be either ET/LT or mergers (see Section~\ref{sec_morph_class}). The dashed line separates the areas where most of the early- 
and late-type galaxies are located (right and left area from the dashed line, respectively). \textit{Top right:} Relation between the smoothness and Abraham concentration index. The dashed line separates the areas where most 
of the early- (on the right; red dots) and late-type (on the left; blue stars) galaxies are located. Relation between the mean surface brightness and 
M$_{20}$ moment of light (\textit{Middle left}), and between the M$_{20}$ moment of light and Abraham concentration index (\textit{Middle right}). Symbols are the same as on the previous two plots. High probability regions of 
finding any of three morphological groups (early-, late-, or compact sources) have been defined in both plots. \textit{Bottom left:} Relation between the M$_{20}$ moment of light and Gini coefficient. See previous plots for 
symbols description. \textit{Bottom right:} Relation between the mean surface brightness and Gini coefficient. See previous plots for symbols description.
\label{fig_morphdiagrams_all}}
\end{figure*}

In order to check the
obtained probabilities, a visual inspection of the images was also performed. It represents a traditional way of classifying galaxies between different morphological types, and it is rather subjective, 
but for bright and extended objects it can produce a high confidence level morphology, and may be a 
helpful tool to probe the reliability of different structural parameters computed with automated methods. To this end, the IRAF/\textit{imexam} tool has been used in order to obtain the isophotal contour diagrams, providing 
information of bulge/disc dominated objects \citep[for more information see][]{povic09a}.
For objects classified as compact and late-type there is a good correspondence between automated and visual classification. Conversely, for objects classified with galSVM as early-type galaxies, having $p_1$ 
probabilities $\ge$\,0.75 there is a good correlation with visual inspection, while in cases of lower $p_1$ probabilities (0.5\,$<$\,$p_1$\,$<$\,0.75), the number of possible interactions starts to grow, as well as the number 
of objects with low S/N ratio, being possible late-type objects. Therefore, in the further analysis only high-probability ($p_1$\,$\ge$\,0.75), early-type objects have been considered in the early-type class. \\
\indent Figure~\ref{fig_morphdiagrams_all}a shows a standard morphological
classification diagram, comparing the asymmetry parameter (A) and Abraham
concentration index (C). This Figure has been used roughly to separate the areas where most of early- (only objects with $p_1$\,$\ge$\,0.75; red dots) and late-type galaxies (all objects with $p_2$\,$>$\,0.5; blue stars) are 
located. The separation is shifted with respect to that defined by 
Abraham et al. (1996; see their Figure 5). This could be expected at for least two reasons: first, a number of early-type spirals (between S0 and Sa galaxies) enter in the early-type 
classification obtained by galSVM , and/or second, different methods have been used in order to obtain concentration and asymmetry indexes. Moreover, as already mentioned, in this work the 
morphological classification was not solely based on this two-parameter diagram, but rather on 8 parameters that have been used by galSVM (see Table~\ref{tab_morph_class_set_sxds}). As 
discussed in more detail in Section~\ref{sec_morph_difficulties}, it seems that a multi -parameter space with non-linear separations is necessary in order to separate early- from 
late-type galaxies accurately. This is one of the main galSVM advantages in comparison to other morphological classification codes. In the same Figure we also show objects having early-type class probabilities  
0.5\,$<$\,$p_1$\,$<$\,0.75 (open diamonds). These objects cover both areas, with a possibility of being either early-, late-type galaxies or interacting systems, as explained above.

\subsection{Difficulties in morphological classification}
\label{sec_morph_difficulties}

- \textbf{\textit{Systematic trend with brightness, size, and redshift.}} The relations between the different morphological parameters and apparent magnitude (in the $i'$ band), 
size (isophotal area), and distance (redshift) have been tested for a sample of SXDS X--ray emitters analysed in this paper, belonging to different morphological types: compact, early- ($p_1$\,$\ge$\,0.75), 
and late-type objects ($p_2$\,$>$\,0.5). Figure~\ref{im_morph_vs_zmagiso_sxds} shows the mentioned relations for the asymmetry and Abraham concentration indexes, Gini coefficient, 
M$_{20}$ moment of light, and mean surface brightness. The mean number of objects and S/N ratio in $i'$ band are represented as well. There is a certain trend of observed morphological parameters with the apparent 
magnitude, size, and distance. 
Parameters related with the galaxy concentration (C, Gini, and M$_{20}$) 
seem to be more affected than the asymmetry index. In other words, fainter, smaller and more distant objects show systematically lower light concentrations. This can be due to detection effects, but aside from this, 
it might also be related to the intrinsic properties of the galaxies: fainter and/or smaller 
objects are intrinsically less concentrated \citep[e.g.,][]{blanton01}. In general, the parameters of the objects detected as compact seem to be less affected with brightness, size, and 
redshift than early- or late-types. On the other hand, early- and late-type galaxies seem to show similar trends. However, as already mentioned above, the code is taking these trends into account in the 
final classification since the same ones are also present in the local training sample.

- \textbf{\textit{Low S/N ratios.}} The bottom panels in Figure~\ref{im_morph_vs_zmagiso_sxds} show how the S/N ratio in $i'$ band changes with brightness, size, and distance. As already mentioned in the previous 
two sections, a low signal-to-noise ratio increases the uncertainty in the morphological classification. Two effects can therefore be observed:
\begin{itemize}
\item As the S/N decreases, the information from the galactic disc can be lost. This results in the possibility that a group of late-type galaxies can be classified as early-type, while early-type galaxies may be detected as 
compact. Low S/N ratio affects all morphological parameters, those related with the asymmetry/smoothness, as well as those related with the galaxy light concentration.
\item Low S/N ratios can affect the bulge information as well. For example, it can be observed how the Abraham concentration index changes as the S/N decreases in a bulge-dominated or compact object. It can be expected that a 
low S/N ratio will smooth out the galaxy profile affecting regions far away from the centre more severely. Thus the galactic radius (used to calculate the concentration index as the ratio between the total flux and the 
flux at 30\% of the radius) will decrease as well. Both F$_{tot}$ and F$_{30}$ fluxes will be affected, but larger impact will correspond to F$_{30}$, which could decrease significantly, leading to lower values of the galaxy 
light concentration. Therefore, bulge-dominated objects with low S/N ratios will have lower concentrations (than the ones with high S/N), placing them in the regions populated by late-type objects. Nevertheless, 
most of these diagrams can be used to define the region where most of the compact objects are located, as shown in Figure~\ref{fig_morphdiagrams_all}.
\end{itemize}

- \textbf{\textit{Number of parameters needed for the morphological classification.}} As already stated, various structural parameters have been obtained with galSVM in order to characterise 
the morphology of the X--ray emitters in the SXDS field. The relationship between them has been studied after obtaining the final morphological classification. It has been seen that for the 
observed sample of objects the combination of any two parameters is not enough to separate between early- and late- type galaxies. The classification in a multi-parameter space with non-linear 
separations is needed, as done with galSVM. However, besides the logA vs. logC diagram \citep[see][]{abraham96}, represented in Figure~\ref{fig_morphdiagrams_all}a, the smoothness and M$_{20}$ moment of light 
parameters related with the Abraham concentration index, and the mean surface brightness related with the M$_{20}$ moment of light make it possible to define the best probability regions of finding compact, early- and late-type objects, as 
shown in Figures~\ref{fig_morphdiagrams_all}b, c, and d, respectively, although the dispersion between the classes is still significant. A linear correlation has also been seen between the four parameters related with 
galaxy concentration (Abraham and Bershady-Conselice concentration indexes, Gini coefficient and M$_{20}$ moment of light), as well as between the mean surface brightness and C and 
Gini (see Figure~\ref{fig_morphdiagrams_all}e, f). However, aside from the M$_{20}$-C diagram, for the observed sample they do not provide a good method for separating early- and late-type morphologies. 

\subsection{Morphological classification: summary}
\label{sec_morf_general}
\vspace{0.3cm}

\indent \indent The catalogue that provides the final morphological 
classification and the list of morphological parameters and other data obtained in this work, 
is described in Appendix A, and is available in the electronic edition of this paper.\\
\indent Having performed a morphological study of X--ray emitters with optical counterparts in the SXDS field, it can be concluded that the sources analysed in this paper are predominantly hosted by 
luminous spheroids and/or bulge-dominated galaxies. At least 55\% of the hosts are compact or E/S0/S0-Sa galaxies. 
Table~\ref{tab_general_morph} summarises the final morphological classification of the selected sample, a subsample of the full X--ray population. Although the selected sample seems to be representative of 
the full X--ray population (see Section~\ref{subsec_sample_selection}), the results presented here should not exclude the possibility that the remaining 73\% for which morphological classification was not possible 
may be diminated by later types. 

\begin{table}[ht!]
\begin{center}
\caption{Summary table: Morphology of X--ray emitters with optical counterparts in the SXDS field
\label{tab_general_morph}}
\small{
\begin{tabular}{l l}
\noalign{\smallskip}
\hline
\textbf{Morphology}&\textbf{Comment}\\
\hline
22\% compact&\\ 
30\% early-type&$p_1$\,$\ge$\,0.75\\ 
23\% early-/late-type or possible interactions&0.5\,$<$\,$p_1$\,$<$\,0.75\\ 
18\% late-type&$p_2$\,$>$\,0.5\\ 
7\% unidentified&\\ 
\hline
\end{tabular}
}
\normalsize
\rm
\end{center}
\end{table}

\begin{figure}[ht!]
\centering
\includegraphics[width=9.5cm,angle=0]{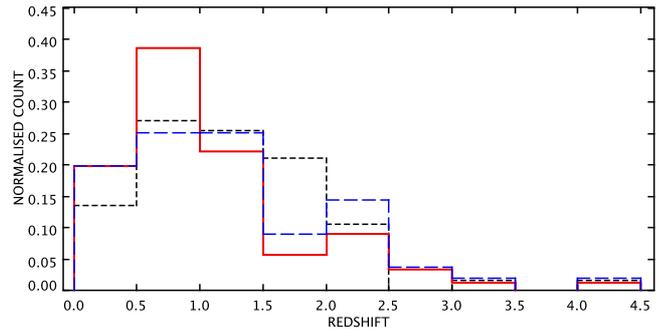}
\protect\caption[ ]{Normalised redshift distributions of compact (short-dashed black line), early- (solid red line), and late-type (long-dashed blue line) sources.  
\label{histo_morph}}
\end{figure}

\indent As will be commented in Section~\ref{subsec_colour_nucty}, most (if not all) of the X--ray emitters analysed in this paper are AGN. Therefore, the results obtained in this work can be compared with 
previous results on host galaxy morphologies of X--ray selected AGN.\\
\indent In general, the morphological classification performed in this work confirms some of the latest findings that indicate that most X--ray detected AGN are hosted by spheroids and/or bulge-dominated 
galaxies \citep[e.g.,][]{grogin05, pierce07, gabor09, povic09a, georgakakis09, griffith10}. \cite{pierce07} used a sample of 94 intermediate-redshift AGN (0.2\,$\leq$\,$z$\,$<$\,1.2), selected 
using \textit{Chandra} X--ray data and \textit{Spizer} MIR data in the Extended Groth Strip (EGS) field. Basing their classification on M$_{20}$ moment of light and Gini coefficient, they found 
that X--ray selected AGN mostly reside in bulge-dominated galaxies (53$^{+11\%}_{-10\%}$), which is in good agreement with our study taking into account both galaxies classified as compact and 
early-type. Our results are also in good agreement with \cite{georgakakis09} who used the high resolution optical data from the Hubble Space Telescope (HST) to study the morphological properties 
of a large sample of X--ray detected AGN. They found that a majority of their objects are bulge-dominated galaxies. Compared with the latest analysis performed by \cite{griffith10} we found a 
lower number of compact sources, but a higher number of early-type bulge-dominated galaxies. \\
\indent Most of the objects morphologically classified have redshifts z\,$<$\,2.0 (although a small fraction of 16\% have photometric redshifts z\,$>$\,2.0), which indicates that 
nuclear activity remains preferentially associated with bulge-dominated galaxies out to substantial look-back times. Figure~\ref{histo_morph} shows the normalised distribution of different morphological types with 
redshift, suggesting that the objects from our sample classified as early-type reach a peak around z\,$\sim$\,0.8, while the maximum distribution of late-type objects peaks at higher 
redshifts, z\,$\sim$\,1.2. Most of the objects detected as compact have been found around z\,$\sim$\,1.0. \\

\indent As shown in \cite{georgakakis09} and references therein, studying the morphology of the AGN host galaxies can be an important way of addressing the issue of the AGN fuelling mechanisms, by 
placing robust limits on the relative contribution of the different mechanisms (e.g., major mergers, minor interactions, internal instabilities) to the accretion history. In the standard view of 
hierarchical structure formation, spheroidal and early-type bulge-dominated galaxies suffered major mergers in their past that destroyed any pre-existing discs to form a bulge 
dominated remnant \citep[e.g.,][]{barnes96, springel05b, hopkins09}. Hence we can conclude that different mechanisms may 
be responsible for putting a galaxy in an active phase. As shown above, at least 50\% of X--ray detected AGN analysed in this work reside in spheroids and/or bulge-dominated systems, 
suggesting that they might have undergone major mergers in their accretion history. However, at least 18\% of the AGN seem to have disc galaxy hosts, showing therefore that minor interactions, internal 
instabilities, and/or some other secular mechanisms could also play an important role in accretion and black hole feeding.

\section{AGN colour-magnitude relations (CMRs)}
\label{sec_colours}

As already mentioned, besides morphology, colours are essential for
studying the properties of AGN host galaxies and their connection with the AGN phenomena. Using a large sample of X--ray selected AGN, we have analysed colour-magnitude diagrams in order to 
study their relationship with other properties of active galaxies. The sample used in this study has two main advantages, when compared to those used in previous works: on the one hand, it is 
much larger, and on the other, it represents one of the deepest optical datasets to date.

\indent To measure the colours of AGN, only objects having
$logX/O\,\ge\,-1$ have been examined, the typical values for active galaxies,
where X--ray fluxes are measured in the 0.5\,-\,4.5\,keV energy range and optical in the R band  \citep[e.g.,][]{fiore03,della04}. Besides the AGN sample
selected in the SXDS field, a sample of AGN selected in the GWS field has also been used \citep{povic09a}. This extended sample has been compared with a 
population of normal galaxies belonging to the CDF-S field \citep{wolf01,wolf04,wolf08}. SXDS optical data are deeper than CDF-S data (and deeper than most (if not all) other ground based photometry optical surveys), 
however only 20\% of active galaxies from our sample have magnitudes above the completeness limit of CDF-S data. Nonetheless, this survey has the information necessary for our comparisons publicly available for a 
large sample of normal galaxies. The CDF-S catalogue contains 50,000 objects; from these, those objects classified by the COMBO-17\footnote{Classifying Objects by Medium-Band Observations - a
spectrophotometric 17-filter survey; http://www.mpia.de/COMBO/combo\_index.html} survey as 'Galaxy' (44,925 sources) have been used in this work as a comparison sample. 
Following \citet{wolf01,wolf04,wolf08}, the selected sample could contain up to a few dozen Seyfert-1 objects, as well as few dozens Seyfert-2 galaxies. 
However, compared with the population of normal galaxies, just a small contribution of these objects is expected to be found \citep{wolf01,wolf04,wolf08}. Therefore, we do not expect that 
such a small fraction of these active galaxies can contaminate significantly 
the sample of non-active galaxies selected and used in this work, thereby changing the derived conclusions. Moreover, considering that
the sample of normal galaxies has been classified by the COMBO-17 survey using all the available spectrophotometric information in 17 filters, it should be good when comparing 
the colours between normal and active galaxies, without introducing a possible selection bias. \\
\\
\indent As noted in Section~\ref{subsec_sample_selection}, in all colour-magnitude diagrams, we only analysed objects with photometric redshifts z\,$\le$\,2 and with errors below 20\%, in order to avoid the 
possibly less reliable measurements. There are 262 objects in total fulfilling these conditions.

\subsection{CMRs in relationship with redshift}
\label{subsec_colour_z}

\begin{figure}[!ht]
\centering
\vspace*{1truecm}
\includegraphics[width=8.4cm,angle=0]{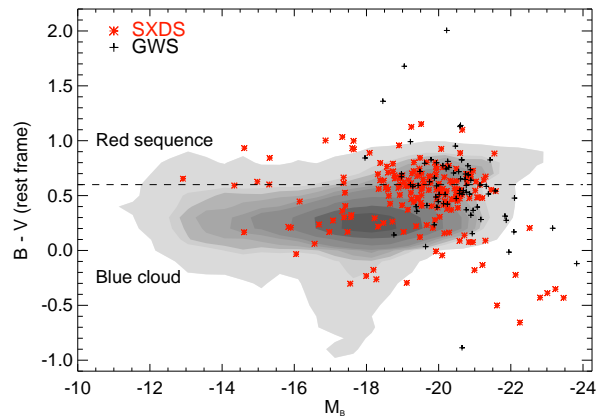}
\caption[ ]{Colour-magnitude diagram showing the relationship between the rest-frame B\,$-$\,V colour and the absolute magnitude in the B band for a sample of AGN in the SXDS (red stars) 
and GWS (black crosses) fields having redshifts  z\,$\le$\,2. The sample of AGN is compared with a sample of normal galaxies from the CDF-S field \citep{wolf01,wolf04,wolf08} represented by contours. 
The gray scales of the contours are scaled to the data, where the darkest and brightest show the highest and the lowest density of the sources, respectively. 
The dashed line shows the \cite{melbourne07} separation between the red sequence (B\,$-$\,V\,$>$\,0.6) and blue cloud (B\,$-$\,V\,$<$\,0.6) galaxies. Median error bars are 0.0083, and 0.0325 for the absolute magnitude in the B band, and 0.0198, and 0.0489 for the B\,$-$\,V rest-frame colour, in the SXDS and GWS fields, respectively.
\label{fig_cmd_allz}}
\end{figure}

\begin{figure*}[!ht]
\centering
\vspace*{1truecm}
\includegraphics[width=14.4cm,angle=0]{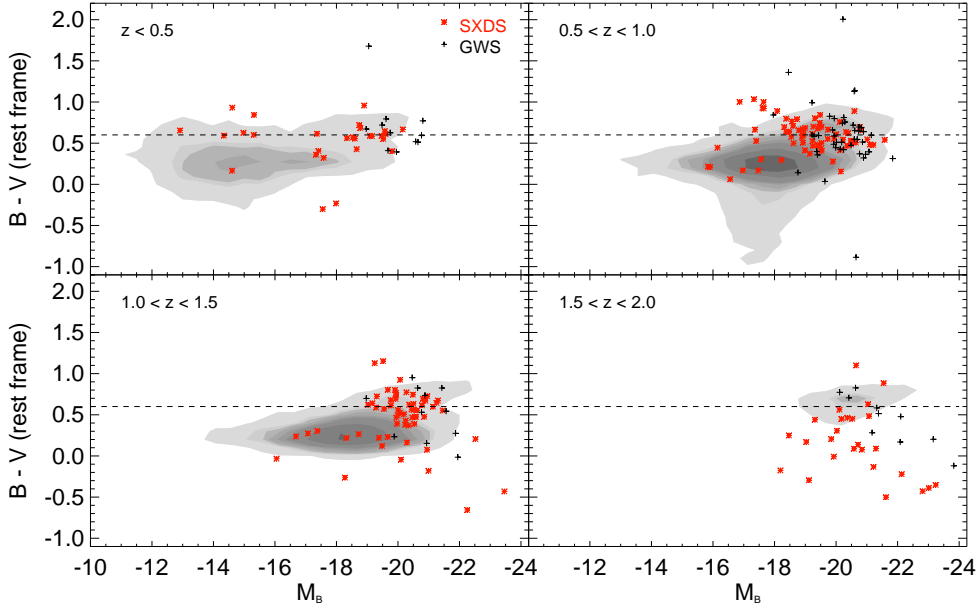}
\caption[ ]{The same colour-magnitude diagram as shown in Figure~\ref{fig_cmd_allz}, but presented in four redshift intervals, from up to down, and from left to right: z\,$\le$\,0.5, 0.5\,$<$\,z\,$\le$\,1.0, 1.0\,$<$\,z\,$\le$\,1.5, and 1.5\,$<$\,z\,$\le$\,2.0. 
\label{fig_cmd_zinter}}
\end{figure*}

Figure~\ref{fig_cmd_allz} shows the CMR between the rest-frame B\,$-$\,V colour and
the absolute magnitude in the B band ($M_B$) for the population of AGN in the SXDS and GWS fields, and for a comparison sample of normal galaxies from the CDF-S field. 
Moreover, this CMR has been analysed in four redshift intervals: z\,$\le$\,0.5, 0.5\,$<$\,z\,$\le$\,1.0, 1.0\,$<$\,z\,$\le$\,1.5, and 1.5\,$<$\,z\,$\le$\,2.0, as shown in 
Figure~\ref{fig_cmd_zinter}. 
Rest-frame colours of normal galaxies have been obtained from the CDF-S catalogue \citep{wolf01,wolf04,wolf08}, while for the sample of AGN the rest-frame colours and the absolute magnitudes have been 
computed from the k-corrected apparent magnitudes. The population of normal galaxies shows the already mentioned bi-modality of colours at all observed redshifts. We plotted on all diagrams 
the \cite{melbourne07} rest-frame colour separation (B\,$-$\,V\,$=$\,0.6), in order to distinguish between galaxies belonging to the red sequence (those with B\,$-$\,V\,$>$\,0.6) and the 
blue cloud (those with B\,$-$\,V\,$<$\,0.6).

As shown in Figures~\ref{fig_cmd_allz} and \ref{fig_cmd_zinter}, we can confirm some of results found in previous works \citep[e.g.,][]{nandra07,georgakakis08,silverman08,schawinski09}: most X--ray selected AGN 
reside in the green valley (the region between the red sequence and blue cloud), at the 
bottom of the red sequence, and at the top of the blue cloud. At least $\approx$\,60\% of the AGN in the SXDS and GWS fields are located in this region of the colour-magnitude diagram, 
having absolute $M_B$ magnitudes in the range -18.0\,$\lesssim$\,$M_B$\,$\lesssim$\,-21.5, and rest-frame colours
in the range 0.3\,$\lesssim$\,B\,$-$\,V\,$\lesssim$\,0.9. If we compare the distributions of rest-frame B\,$-$\,V colours for red and blue AGN and for normal galaxies, it can be 
seen that the distribution of red AGN peaks towards bluer colours (B\,$-$\,V\,$\approx$\,0.62) while the distribution of blue AGN peaks toward redder colours (B\,$-$\,V\,$\approx$\,0.57), in comparison with the sample of 
red and blue normal 
galaxies (B\,$-$\,V\,$\approx$\,0.7 and $\approx$\,0.3, respectively). As shown in Figure~\ref{fig_cmd_zinter}, AGN start to populate the green valley at higher redshifts, z\,$>$\,0.5. Analysing the number of sources in 
redshift bins with a width
of 0.1, the maximum number of AGN has been found at z\,$\approx$\,0.9. \\
\indent Besides the green valley where most AGN reside, a number of very luminous ($M_B$\,$\lesssim$\,-21.5), blue
 sources (rest-frame B\,$-$\,V\,$\lesssim$\,0.2) lie outside the region
 covered by the normal galaxy population (Figure~\ref{fig_cmd_allz}). They are all very
luminous X--ray sources, having luminosities L$_X$\,$\ge$\,10$^{44}$\,erg\,s$^{-1}$ in the 0.5\,-\,7.0\,keV energy range.
Moreover, practically all of these sources have been classified as compact (see Section~\ref{subsec_colour_morph} for more details). This region corresponds to QSO objects. Around 10\% of the total sample of the AGN 
studied in this work reside in this area. They start to populate the colour-magnitude diagrams at higher redshifts (z\,$>$\,1), as shown in Figure~\ref{fig_cmd_zinter}.\\
\indent In comparison with previous works, we detected a much larger population of
low-luminosity AGN ($M_B$\,$\gtrsim$\,-18.0),
most of them at redshifts $<$\,1.0, being located in both the blue cloud and red sequence.
Around 15\% of AGN from the whole sample have been found
to lie in this region. These AGN can occupy a wide range
of rest-frame colours (-0.4\,$\lesssim$\,B\,$-$\,V\,$\lesssim$\,1.1),
with most of them having X--ray luminosities above 10$^{43}$\,erg\,s$^{-1}$ in the 0.5\,-\,7.0\,keV energy range. All these objects belong to the SXDS field and have been detected for the first time thanks to 
the large depth of the optical data. 

\begin{figure*}[ht!]
\centering
\begin{minipage}[c]{.49\textwidth}
\includegraphics[width=8.4cm,angle=0]{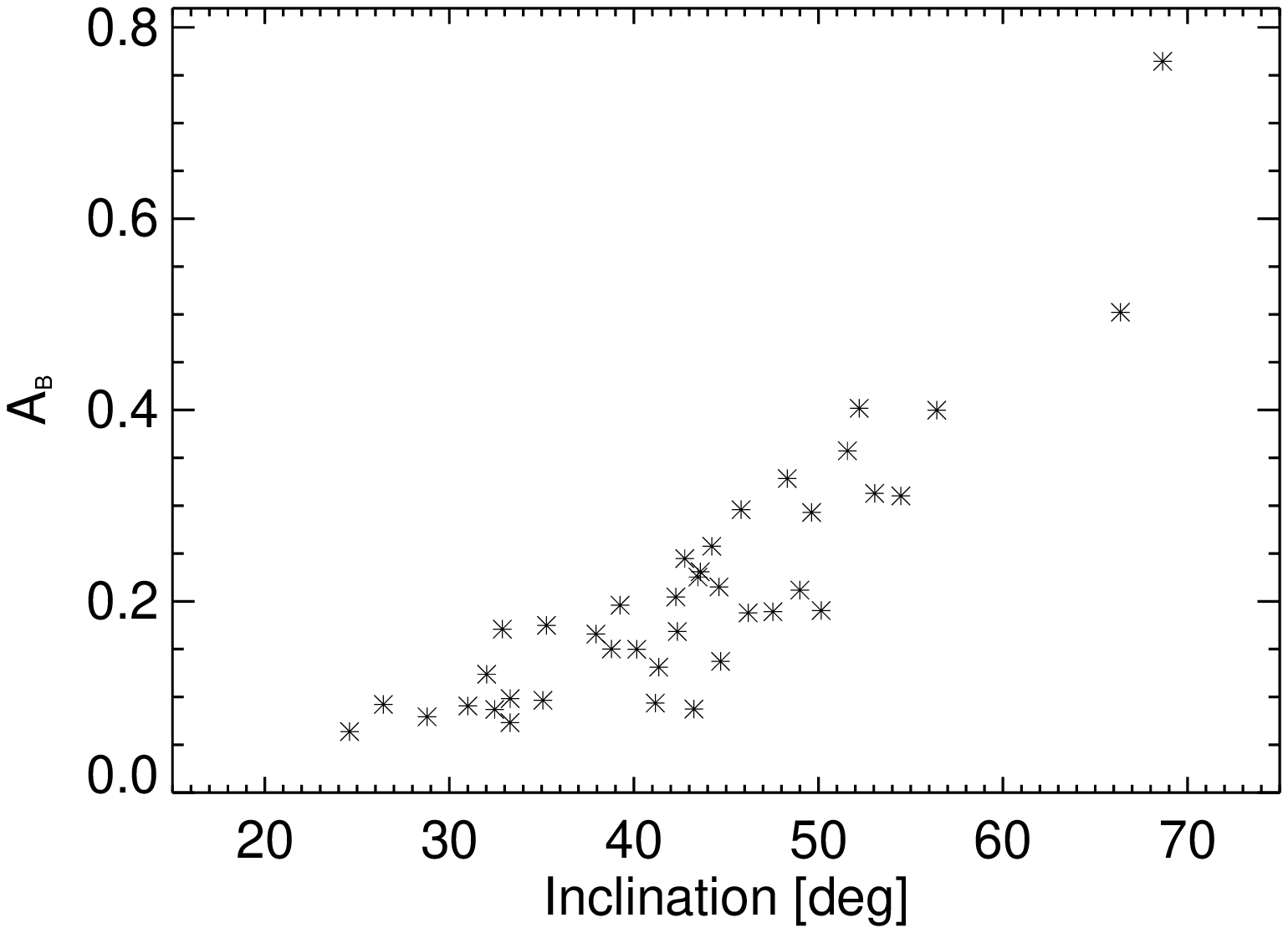}
\centering
\end{minipage}
\begin{minipage}[c]{.49\textwidth}
\includegraphics[width=8.0cm,angle=0]{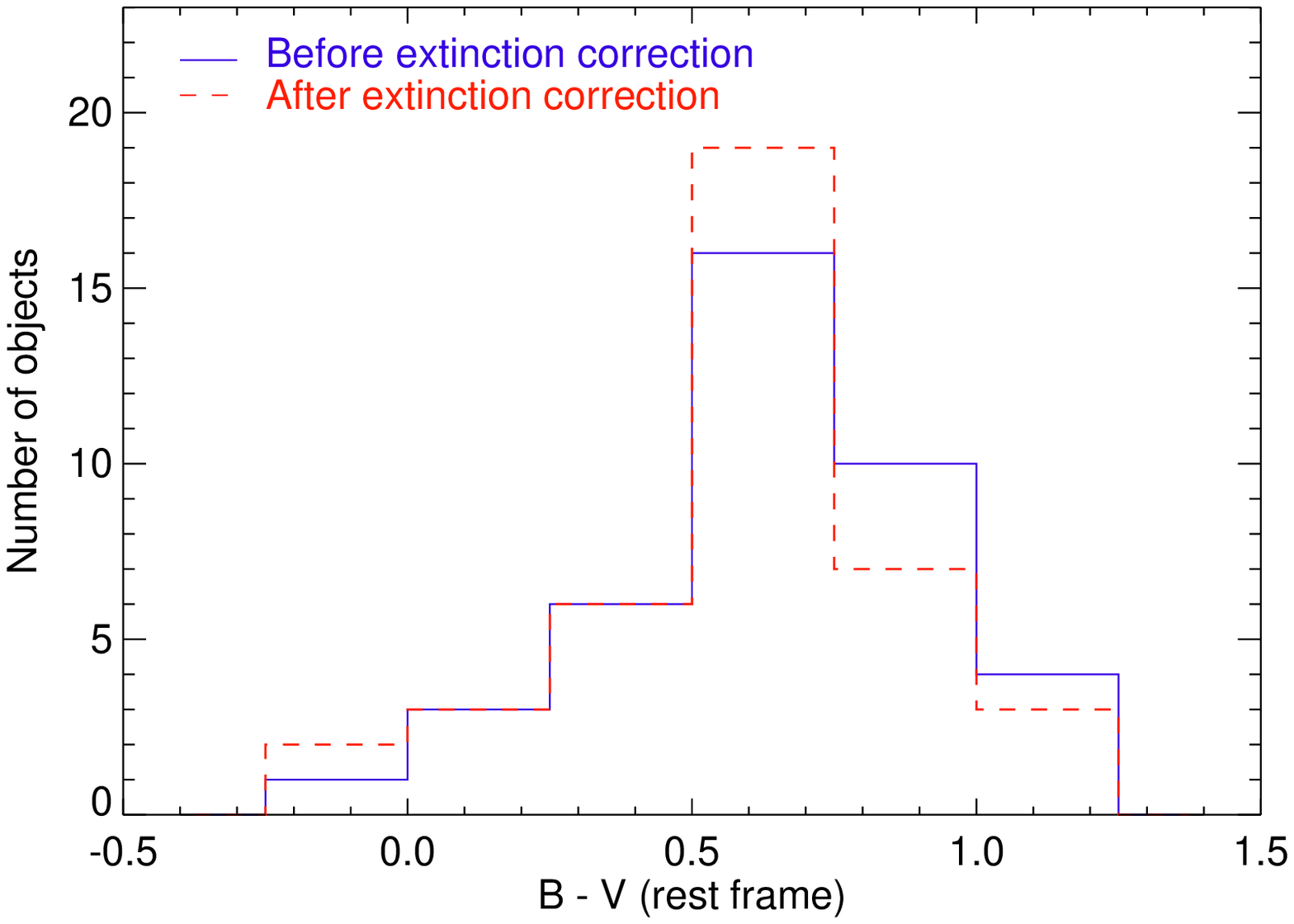}
\centering
\end{minipage}
\caption[ ]{\textit{Left:} Relationship between the extinction $A_B$ measured in B band and inclination for late-type AGN host galaxies in the SXDS field.\textit{Right:} Histogram showing the 
distribution of B\,$-$\,V colour before (solid blue line) and after (dashed red line) extinction correction for late-type AGN host galaxies selected in the SXDS field.
\label{fig_incextAb_histMbMvextnoext}}
\end{figure*}

As already shown, it seems that the colours of X--ray detected active galaxies differ from those of non-active galaxies, in all observed redshift intervals. Normal galaxies show clear bi-modality at all 
redshifts, belonging to the red sequence or blue cloud, while no colour bi-modality has been found for AGN . Usually galaxies residing in the green valley are considered to be transition sources, 
transiting from blue, star-forming galaxies to red, passively-evolving early-type objects. Therefore, since most AGN have been found in this region, it has been suggested 
in several previous works that there is a relationship between the mentioned transition and the AGN activity, and that AGN feedback might be responsible for quenching the star formation in 
blue cloud galaxies, moving them to earlier types \citep[e.g.,][]{nandra07,georgakakis08,silverman08,treister09,springel05,schawinski06, hasinger08}.\\
\indent However, there are a few aspects that should be considered when comparing colours of normal and active galaxies:\\
\indent  - \textbf{\textit{Completeness limit of the comparison galaxies.}}
We checked the completeness limits of both the AGN and control samples, finding that $\sim$\,20\% of AGN ($\approx$\,22\% and $\approx$\,17\% in the B band and V band, respectively) 
have magnitudes above the completeness limit of the comparison non-active galaxies, which is insufficient population for changing the general picture of AGN distribution in the 
colour-magnitude diagrams.\\
\indent  - \textbf{\textit{AGN contribution.}} The AGN contribution to the host galaxy
emission has not been quantified in this work. It has been analysed by
other authors, suggesting that the galaxies hosting AGN with
low X--ray luminosities (logL$_X$\,$\lesssim$\,44), being mostly
located in the green valley, should be moved to even redder colours after the AGN contribution is eliminated and the total galaxy flux corrected 
\citep{kauffman07,nandra07,silverman08}. Recently, using HST images, \cite{cardamone10} studied how the light from the central nuclear source affects the colours of moderate luminosity 
AGN host galaxies in the GOODS survey. They found that the integrated optical galaxy light is dominated by host emission, and that optical colours are not significantly affected by AGN emission. 
Moreover, \cite{pierce10} found that only in some $<$\,10\% (when the AGN is very luminous, unobscured, and/or compact) X--ray selected AGN can affect significantly integrated optical and UV colours; 
otherwise, the AGN contribution to the integrated optical/UV colours is not significant. \\
\indent  - \textbf{\textit{Dust reddening.}}
In several previous works, it has been found that many blue, star-forming galaxies being reddened by dust have colours that would place them in the green valley. 
This could mean that many green valley galaxies are not transition sources, evolving from blue to passively-evolving red galaxies, but rather blue cloud sources being affected by dust. Moreover, 
a possible selection effect could be introduced, observing AGN less affected by extinction, since the most reddened AGN might have magnitudes below the detection limits.    
Using NIR colours to separate between the red early-type galaxies and galaxies being reddened by dust (at 0.8\,$\le$\,z\,$\le$\,1.2), \cite{cardamone10} found that $\sim$\,25\% and $\sim$\,75\% of 
AGN belonging to the red sequence and green valley, respectively, actually have colours typical of young stellar populations being reddened by dust, and that their dust-corrected optical colours 
are blue and similar to those of star-forming galaxies. However, using dust-corrected optical colours, \cite{xue10} found a very weak colour bi-modality for AGN host galaxies at redshifts 
z\,$\le$\,1.0, and no evidence at higher redshifts, suggesting that even after dust-correction active galaxies still appear redder than non-active galaxies, at all redshifts. We have studied the 
effect of dust-reddening in our sample of late-type AGN selected in the SXDS field. In order to obtain the extinction in both B and V bands, and the extinction corrected rest-frame 
B\,$-$\,V colour, we used the same corrections as in \cite{FernandezLorenzo10}. As noted in that paper, using the \cite{tully98} method the extinction $A_B$ in the B band can be measured as a 
function of the inclination $i$, through major-to-minor axis ratio $a/b$: $A_B$\,=\,$\gamma_B$\,log($a/b$), with $\gamma_B$\,=\,-0.35\,(15.31\,+\,$M_B$). These formulae have been derived using 
a sample of spiral galaxies and the Vega magnitudes. Therefore, we are dealing with Vega magnitudes when analysing the extinction correction, studying only galaxies classified as late-types. The 
major and minor axes have been obtained by SExtractor. As noted by \cite{FernandezLorenzo10}, using \cite{calzetti00} the extinction in the V band can be calculated as $A_V$\,=\,0.8\,$A_B$. 
Figure~\ref{fig_incextAb_histMbMvextnoext} (left panel) shows the relationship between the inclination and the obtained extinction in the B band. After applying the extinction corrections, we did not find any significant 
difference in the colour distribution of late-type AGN. 
Figure~\ref{fig_incextAb_histMbMvextnoext} (right panel) shows the comparison for the B\,$-$\,V colour before and after applying the extinction correction. Moreover, applying the Kolmogorov-Smirnov analysis, 
we obtained a significance level of 0.72, showing that the cumulative distribution function of B\,$-$\,V colour is not significantly different before and after applying the extinction correction. 
This would mean that even after the extinction correction, the late-type galaxies, which should be more affected by dust reddening as compared to early-types, will mostly stay placed in the 
green valley. However, the extinction correction applied here is only a coarse one, since the used equations have been derived for the local universe, and no corrections related with the 
possible evolution of dust content with the redshift have been applied.

\begin{figure}[!ht]
\centering
\vspace*{1truecm}
\includegraphics[width=8.4cm,angle=0]{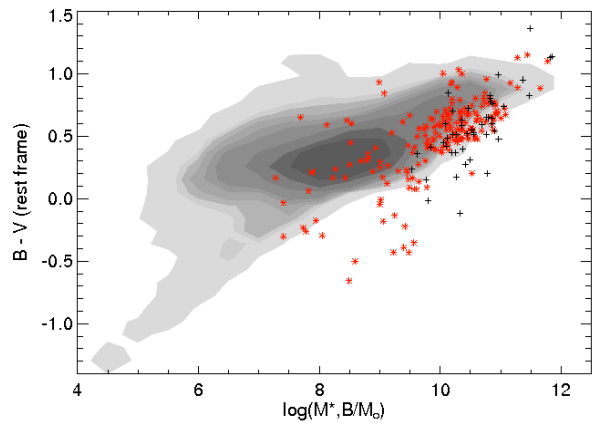}
\caption[ ]{Relation between the rest-frame B\,$-$\,V colour and the stellar mass in the B band for a sample of AGN in the SXDS (red stars) and GWS (black crosses) fields having redshifts  z\,$\le$\,2. See 
Figure~\ref{fig_cmd_allz} for a description of a control CDFS sample represented with grey contours.
\label{fig_bv_stmassB}}
\end{figure}

\indent Conversely, other results show that when comparing samples of active and normal galaxies with similar stellar masses and being at similar redshifts, AGN host galaxies do not show a 
strong evidence for either quenched or elevated star formation \citep[e.g.,][]{alonso08}. Moreover, studying the CMRs of X--ray selected active and non-active galaxies at high redshifts 
z\,$\approx$\,1\,-\,4, \cite{xue10} found a colour bi-modality in non-active galaxies that is absent in AGN hosts. This holds up to z\,$\approx$\,3 (in good agreement with the results mentioned above). Yet 
using stellar mass-selected samples they found that the difference in the colour distribution between active and non-active galaxies disappears, and that mass-selected AGN 
hosts have the same bi-modal distribution in the CMRs as non-active galaxies, up to z\,$\approx$\,2\,-\,3. The authors suggested that AGN preferentially reside in massive galaxies that 
normally tend to have redder colours. However, these results are in contrast with the ones obtained by \cite{schawinski10} 
in the nearby universe, where the colours of mass-selected AGN still peak in the green valley. \\
\indent We measured the stellar masses of our active galaxies (from both the SXDS and Groth fields) and of the control sample using the same procedure explained in \cite{xue10}. 
Figure~\ref{fig_bv_stmassB} shows the relation between the stellar mass in B band and rest-frame B\,-\,V colour. For a sample of AGN studied in this work, even when mass-matched with a sample of inactive galaxies, 
we do not find a clear colour bi-modality. Around 60\% of AGN from our sample reside between the red sequence and blue cloud, having stellar masses between 10$^{9.7}$ and 10$^{11}$\,M$_{\bigodot}$. This is a similar 
population to that found residing in the green valley of colour-magnitude diagrams (see Figure~\ref{fig_cmd_allz}). However, in comparison with \cite{xue10}, we are dealing with different stellar mass populations, that have 
a lower fraction of galaxies at redshifts between 1.0 and 2.0, and without a population of the most massive galaxies (above 10$^{11}$\,M$_{\bigodot}$) they found residing in the red sequence.

\subsection{CMRs in relationship with morphology}
\label{subsec_colour_morph}

As mentioned above, CMRs have been used in previous studies as a tool to analyse the role of AGN activity and AGN feedback in galaxy evolution \citep[e.g.,][]{nandra07,georgakakis08,silverman08,schawinski09}. 
In previous works this type of analysis has usually been performed 
using a whole population of AGN. In order to understand the role of AGN in galaxy evolution, it is necessary to study the colour-magnitude diagrams for different morphological types (instead of 
the whole population), at both low and high redshifts. This is one of the first works where CMRs are studied in relationship with morphology using a high-redshift (up to z\,=\,2) magnitude limited AGN sample. 

\begin{figure*}[ht!]
\centering
\begin{minipage}[c]{.51\textwidth}
\includegraphics[width=8.4cm,angle=0]{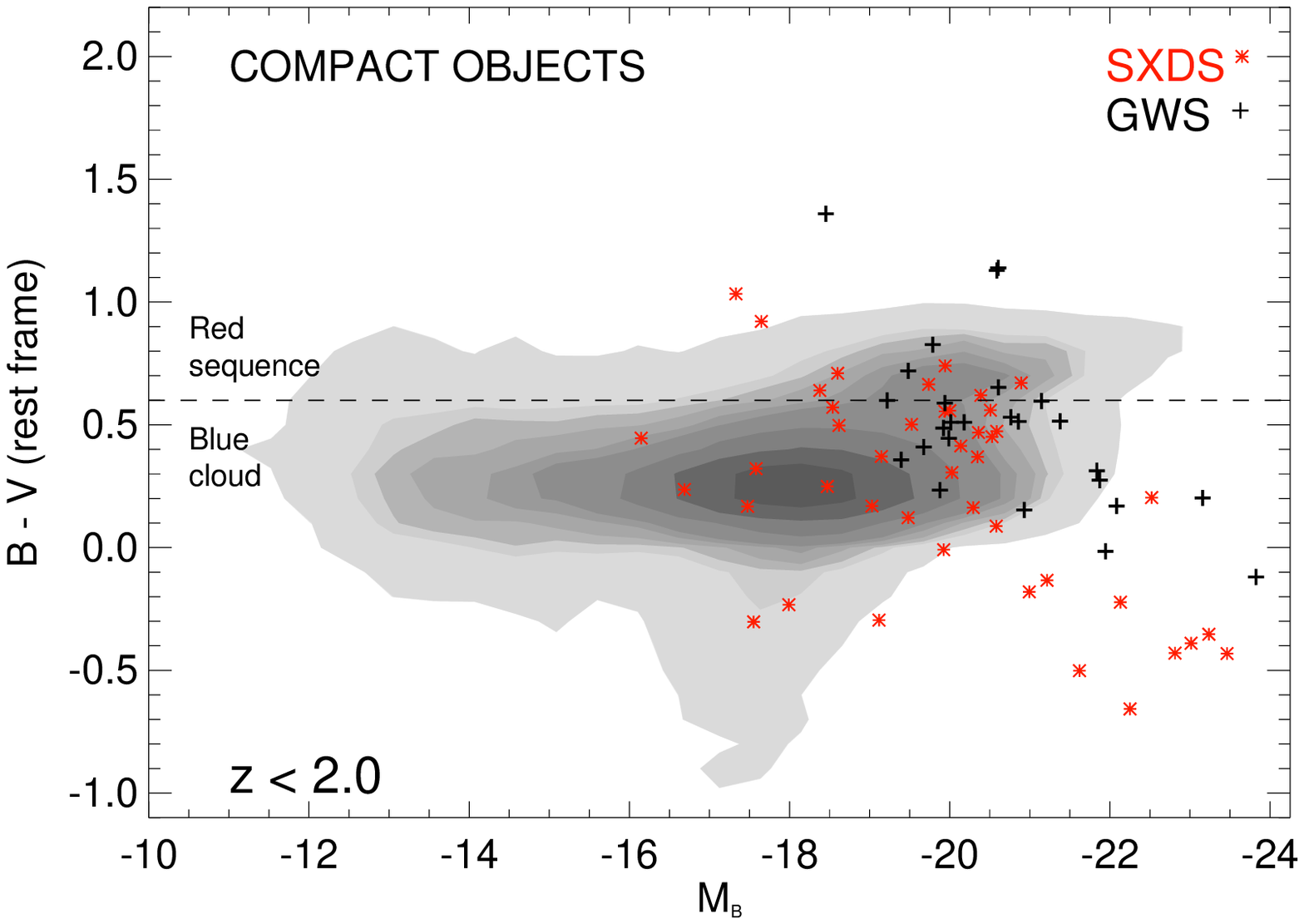}
\centering
\end{minipage}
\begin{minipage}[c]{.49\textwidth}
\includegraphics[width=8.4cm,angle=0]{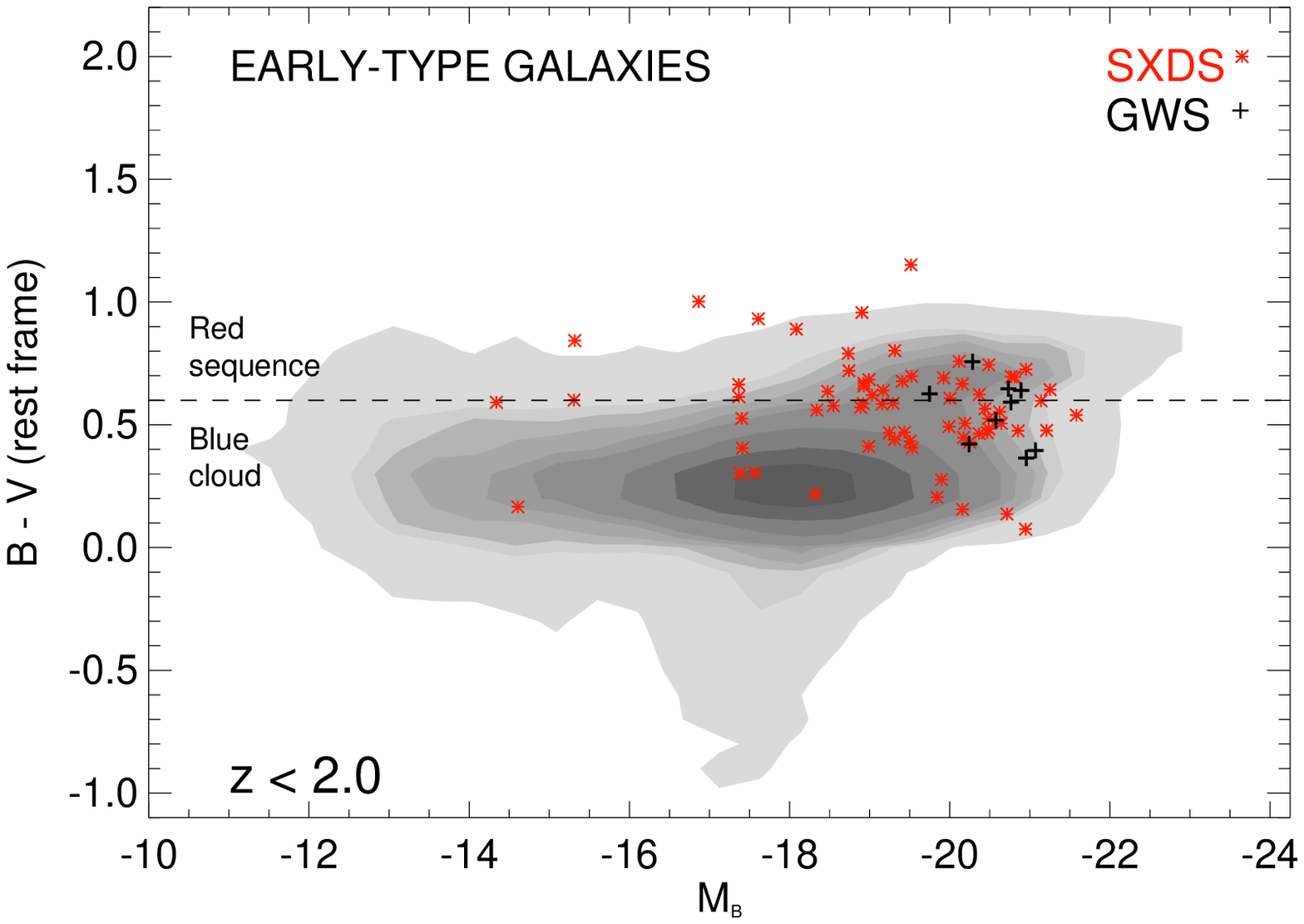}
\centering
\end{minipage}
\begin{minipage}[c]{.49\textwidth}
\includegraphics[width=8.4cm,angle=0]{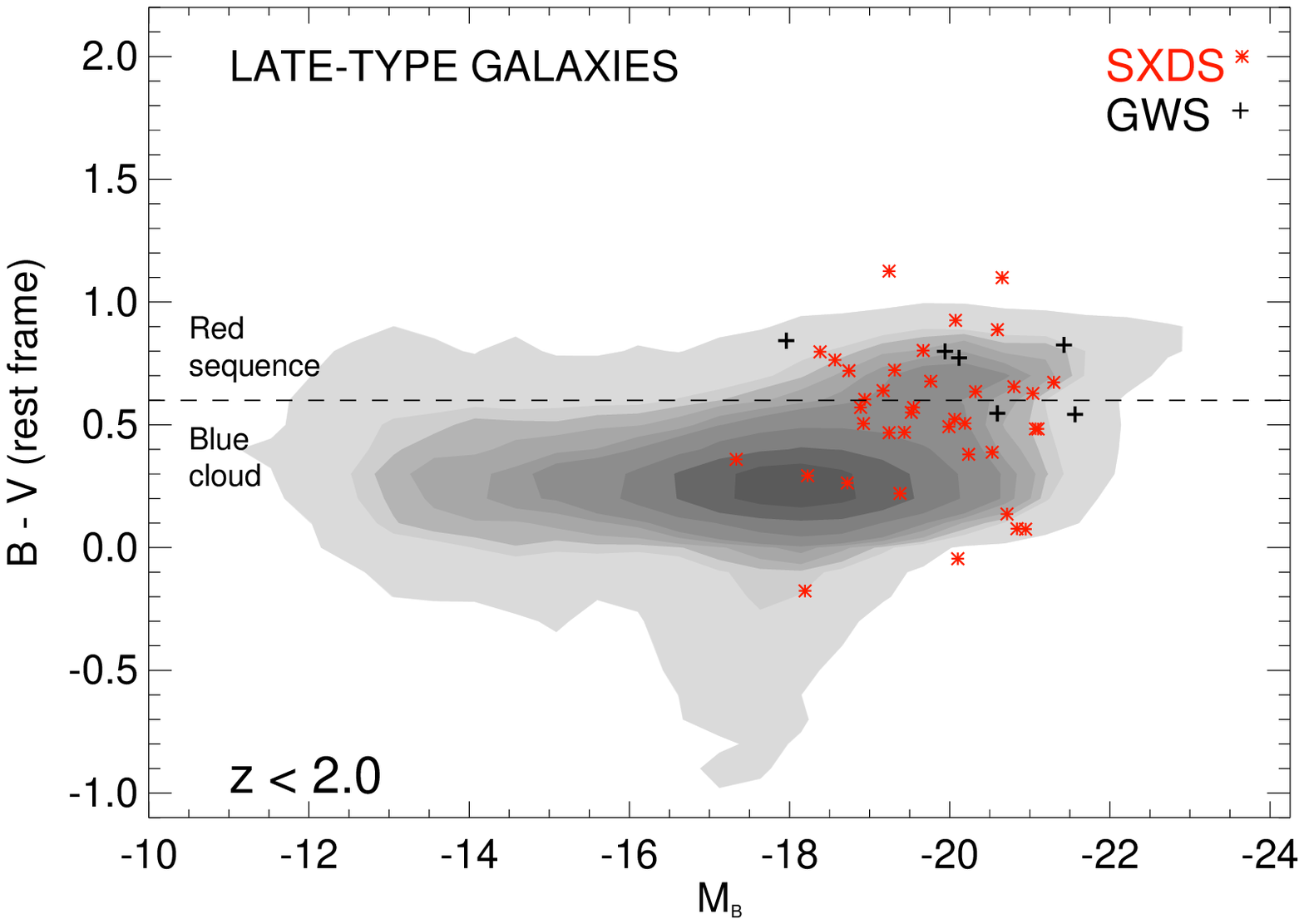}
\centering
\end{minipage}
\caption[ ]{Colour-magnitude diagram showing the relationship between the rest-frame B\,$-$\,V colour and the absolute magnitude in the B band for different morphological types, 
using a sample of X--ray selected AGN in the SXDS (red stars) and GWS (black crosses) fields. All AGN having redshifts z\,$\le $\,2.0 are presented, compact (top), early- (bottom left) and 
late-type (bottom right) galaxy hosts. Morphological classification of active galaxies in the 
SXDS field is described in Section~\ref{sec_morph}, while morphological classification of active galaxies in the GWS field is described in \cite{povic09a}. The sample of AGN is compared with the 
sample of normal galaxies in the CDF-S field \citep{wolf01,wolf04,wolf08} represented with contours. 
Grey scales of the contours are scaled to the data, where the darkest and brightest show the highest and the lowest density of the sources, respectively. 
The dashed line shows the \cite{melbourne07} separation between the galaxies belonging to the red sequence (B\,$-$\,V\,$>$\,0.6) or to the blue cloud (B\,$-$\,V\,$<$\,0.6). 
\label{fig_cmd_morph_allz}}
\end{figure*}

Figure~\ref{fig_cmd_morph_allz} shows the CMRs of X--ray selected AGN in the SXDS and GWS fields, taking into account the morphological classification of sources. Colour-magnitude diagrams 
are presented for compact sources, and for two main morphological types, early- and late-type AGN host galaxies. All AGN with redshifts z\,$\le$\,2.0 are shown. The morphological 
classification of active galaxies belonging to the SXDS field has been described in Section~\ref{sec_morph}. For the selection and morphological classification of AGN in the GWS field see \cite{povic09a}. 
Groups classified as I and II in \cite{povic09a} have been presented here as early-, and groups III and IV as late-type galaxies. As in Figures~\ref{fig_cmd_allz} and \ref{fig_cmd_zinter}, for each 
morphological type the distribution of active galaxies has been compared with the distribution of the global population of normal galaxies from the CDF-S field.

As can be seen in Figure~\ref{fig_cmd_morph_allz}, $>$\,85\% of sources classified as compact have blue colours, with many of them residing in the green valley. A significant number of these sources 
are QSOs, but AGN with faint hosts, and spheroidal galaxies are also presented. Around 25\% of the compact sources reside in the region devoid of normal galaxies, with very blue colours and 
high luminosities typical of high redshift QSOs.\\
\indent On the other hand, no clear separation has been found between early- and late-type galaxy hosts on the colour-magnitude diagrams. Both types seem to follow a similar distribution, being 
again mostly located in the green valley, at the bottom of the red sequence, and at the top of the blue cloud. Therefore, a sample of AGN analysed in this work does not show a clear relationship between the colours and 
morphology. On the other side, colours of active galaxies can be influenced by AGN and/or dust obscuration, moving the 
early-/late-type sources toward the blue cloud/red sequence, respectively. However, as already noted in Section~\ref{subsec_colour_z}, we still need more work in order to quantify these contributions.

\indent As already mentioned above, this is one of the first works in which AGN colour-magnitude relations have been studied according to morphology, for AGN at z\,$\le$\,2. Only recently, \cite{schawinski10} studied 
colour-stellar mass diagrams (instead of colour-magnitude relations) but in the nearby universe (z\,$<$\,0.05) for early- and late-type AGN host galaxies. On these diagrams they did not find a relationship between 
colours and morphology, either, again finding the largest population of AGN in the green valley, with early-type AGN hosts peaking below the red sequence, while the population of AGN in late-type galaxies peaks in 
the green valley, above the blue cloud population of similar masses. Moreover, \cite{mainieri11} studied the colours and morphology of Type-2 QSO at redshifts between 0.8 and 3 in the XMM-COSMOS field, and they 
found that most of their objects have rest-frame colours in the green valley, pointing out that this is the effect of a luminosity-selected rather than a mass selected sample. More than 80\% of their objects have stellar 
masses above 10$^{10}$\,M$_{\bigodot}$, mostly showing bulge-dominated morphologies, and weak signs of recent mergers or discs.  \\

\subsection{CMRs in relationship with X--ray type}
\label{subsec_colour_nucty}
 
\begin{figure}[!ht]
\centering
\vspace*{1truecm}
\includegraphics[width=8.0cm,angle=0]{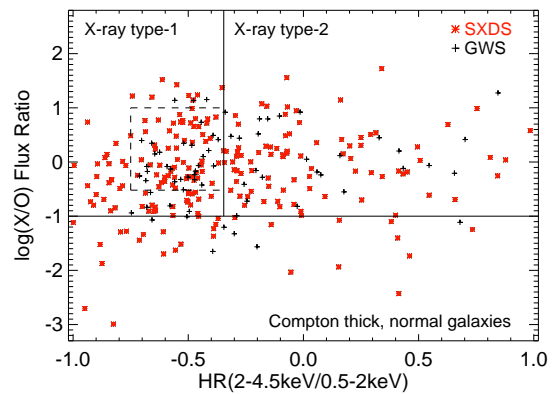}
\caption[ ]{Relationship between the X/O flux ratio and HR$(2-4.5keV/0.5-2keV)$ hardness ratio, for X--ray emitters with optical counterparts in the SXDS (red stars) and 
GWS (black crosses) fields. Solid lines separate X--ray type-1 (unobscured) and X--ray type-2 (obscured) regions and the area with X/O\,$<$\,0.1, where Compton thick AGN, normal galaxies 
and stars can be found \citep[e.g.][]{fiore03}. The dashed line box shows the limits obtained by \cite{della04} where $\sim$\,85\% of their spectroscopically identified AGN with broad emission 
lines have been found. In this work, $\sim$\,51\% of the X--ray type-1 sources are located inside this box.
\label{fig_xohr}}
\end{figure}

In order to study the colour-magnitude diagrams in relation with X--ray obscuration, we have performed a coarse nuclear type classification based on a diagnostic diagram
relating the X/O flux ratio and hardness ratio HR$(2-4.5keV/0.5-2keV)$. The X/O flux ratio has been shown to segregate efficiently between active and non-active/Compton-thick galaxies 
\citep{alexander01,fiore03,civano07}, while the hardness ratio HR$(2-4.5keV/0.5-2keV)$ is very sensitive to absorption, and thus capable of disentangling X--ray type-1 (unobscured or unabsorbed) from X--ray 
type-2 (obscured or absorbed) AGN \citep{mainieri02,della04,perola04,caccianiga04,dwelly05,hasinger08}. Full details on the methodology applied are given in
\citet{povic09a}. According to this, all objects having X/O\,$>$\,0.1 and 
HR$(2-4.5keV/0.5-2keV)$\,$<$\,-\,0.35 have been classified as X--ray type-1 AGN,
while those with X/O\,$>$\,0.1 and HR$(2-4.5keV/0.5-2keV)$\,$>$\,-\,0.35 have been catalogued as type-2
(obscured) nuclei. Those with X/O\,$<$\,0.1 have been classified as
Compton-thick/non-active galaxies. Figure~\ref{fig_xohr} shows the described diagram of X--ray type classification. 52\% and 39\% of the total number of sources with optical
counterparts have been classified as X--ray type-1 and type-2 AGN, respectively. The dashed--line box indicates the locus of the X--ray type-1 region according to Della Ceca et al. (2004). 
This box has been considered as the 'highest probability' region for finding X--ray unobscured AGN, and is populated by 51\% of the X--ray type-1 objects. The remaining
10\% are found in the region typical of Compton-thick/non-active galaxies.

Using the complete sample of objects, we have studied the relationship between the B\,$-$\,V rest-frame colour and the HR$(2-4.5keV/0.5-2keV)$ hardness ratio. 
Both X--ray type-1 and X--ray type-2 sources span the same range of B\,$-$\,V colour, and apparently 
there is no relationship between the observed optical and X--ray colours when the complete population of AGN is considered. However, performing the Kolmogorov-Smirnov analysis produces the result that 
the two distributions might be different, although the evidence is not conclusive, having a probability factor of 0.3.

\begin{figure}[ht!]
\centering
\begin{minipage}[c]{.51\textwidth}
\includegraphics[width=8.4cm,angle=0]{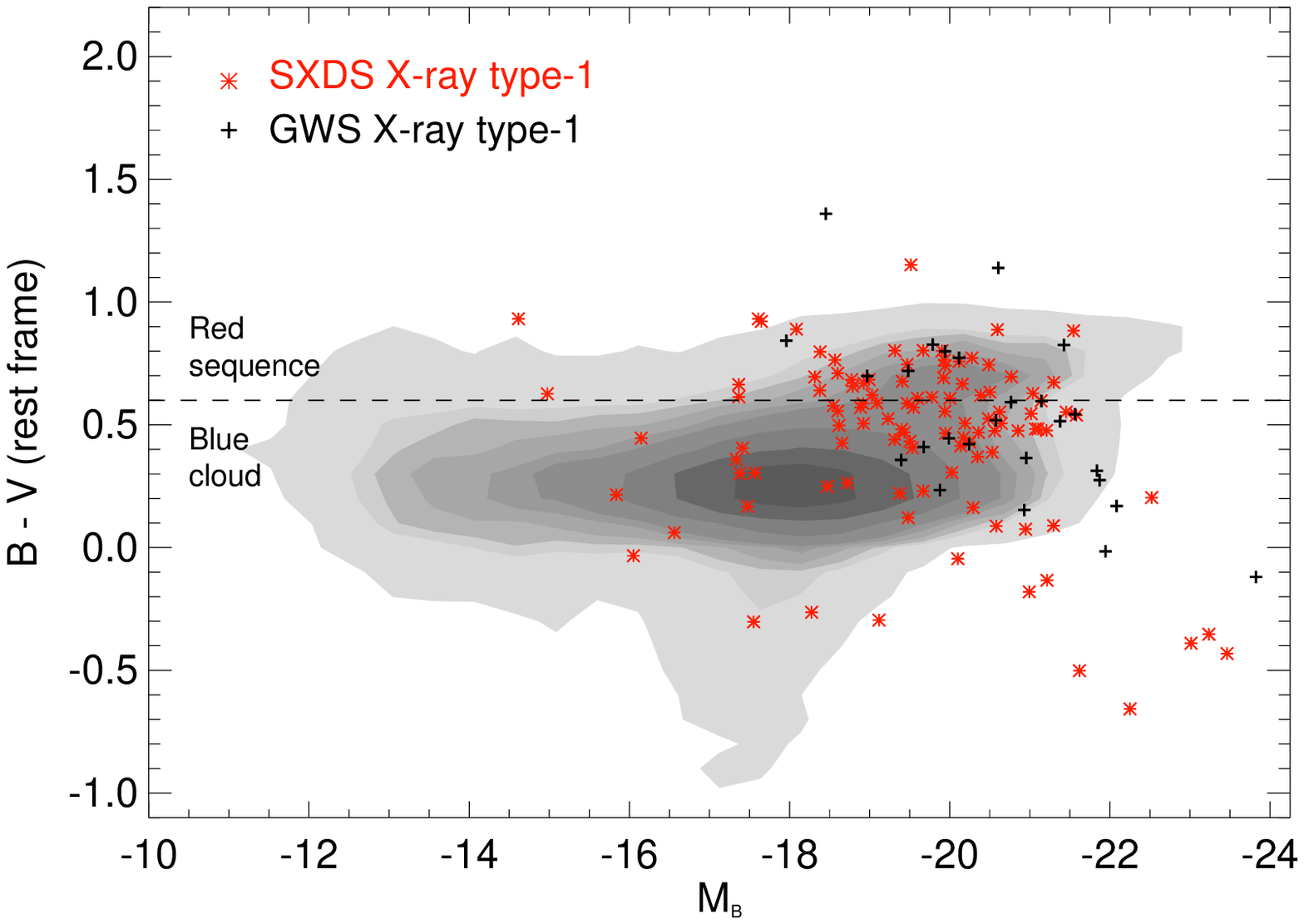}
\centering
\end{minipage}
\begin{minipage}[c]{.51\textwidth}
\includegraphics[width=8.4cm,angle=0]{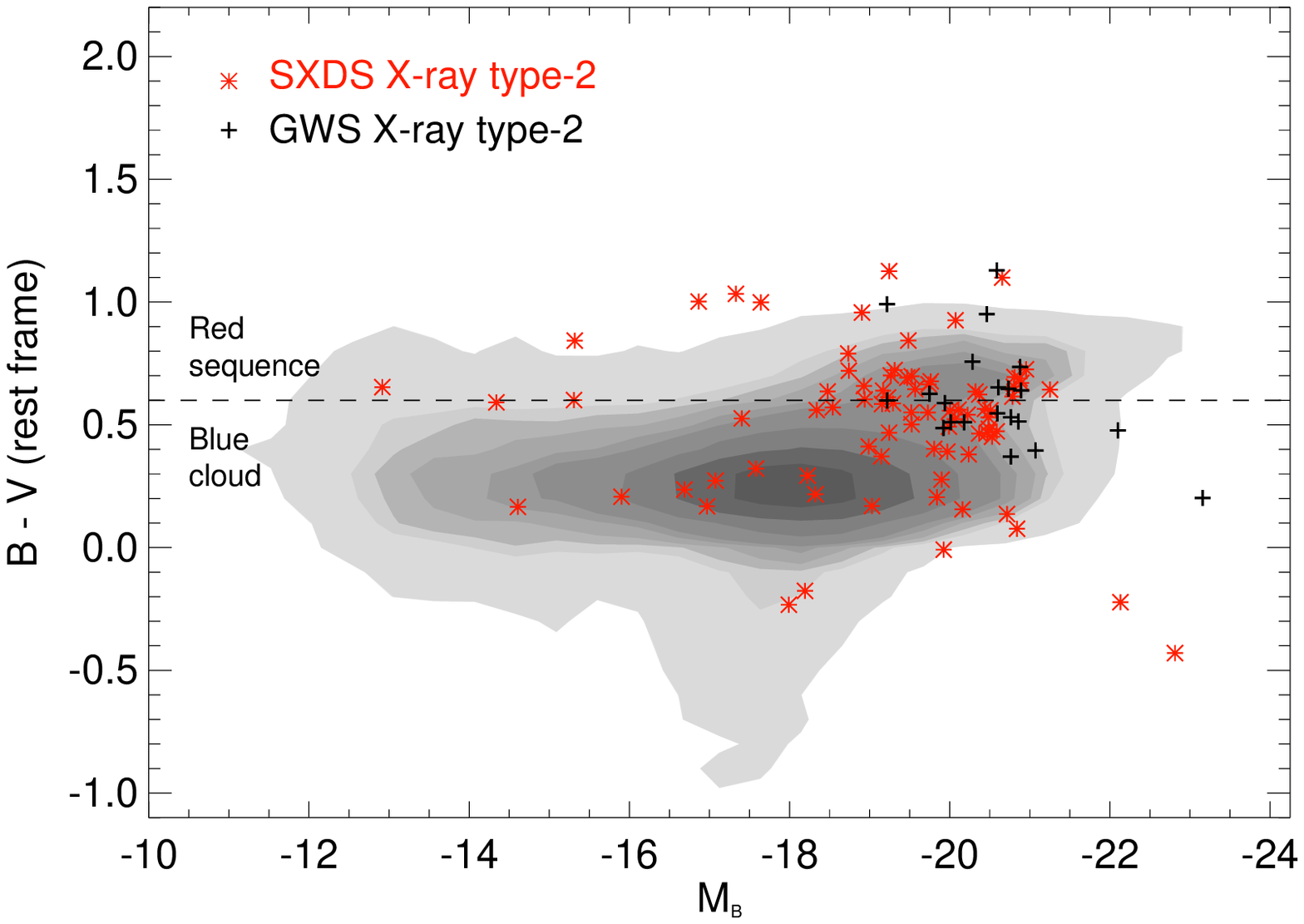}
\centering
\end{minipage}
\caption[ ]{Colour-magnitude diagram showing the relation between the rest-frame B\,$-$\,V colour and the absolute magnitude in the B band for X--ray type-1 (unobscured; \textit{top}) and 
X--ray type-2 (obscured; \textit{bottom}) AGN, selected in the SXDS (red stars) and GWS (black crosses) fields. All sources with redshifts z\,$\le$\,2.0 have been represented. AGN sample has been compared with 
the sample of normal galaxies selected in the CDF-S field \citep{wolf01,wolf04,wolf08} represented with contours. 
The grey scales of the contours are scaled to the data, where the darkest and brightest show the highest and the lowest density of the sources, respectively. 
The dashed line shows the \cite{melbourne07} separation between the galaxies belonging to 
the red sequence (B\,$-$\,V\,$>$\,0.6) or to the blue cloud (B\,$-$\,V\,$<$\,0.6).
\label{fig_cmd_nts}}
\end{figure}

Figure~\ref{fig_cmd_nts} shows the CMRs in relation with the X--ray obscuration. X--ray type-1 and type-2 sources have been represented from both the SXDS and GWS fields, and compared 
with the distribution of the complete sample of normal galaxies from the CDF-S field. As can be seen there is no clear separation between X--ray unobscured and obscured sources in the 
colour-magnitude diagram. We have studied this diagram in four redshift intervals (same intervals as in Figure~\ref{fig_cmd_zinter}), and have found no clear separation between X--ray type-1 and type-2 sources in any of them.
We studied the X--ray obscuration for each morphological type of our analysed sample. Almost all active galaxies (95\%) detected as compact are unobscured sources in X--rays, with most of them (more than 80\%) 
having blue B\,$-$\,V rest-frame colours. 
High luminosity QSO sources, being placed in the region devoid of normal galaxies, are found to be the most unobscured sources, having HR$(2-4.5keV/0.5-2keV)$\,$<$\,-0.5, 
while compared with them, compact sources belonging to the green valley and the blue cloud have lower X--ray obscuration (-0.5\,$\leq$\,HR$(2-4.5keV/0.5-2keV)$\,$\leq$\,-0.36). 
On the other hand, when considering early- and late-type active galaxies, no relation has been found between the morphology and X--ray obscuration. Early- and late-type AGN in our sample hosts present a mixture of 
both X--ray unobscured and obscured sources. For both morphological types, $\approx$\,55\% of 
sources are unobscured, while the remaining $\approx$\,45\% are obscured. Moreover, there is no significant relationship between the X--ray obscuration and optical B\,$-$\,V 
rest-frame colour for both morphological types. 55\% and 50\% of unobscured and obscured early-type AGN are found to have blue optical colours, respectively, with very similar values 
for late-type hosts (58\% and 50\%). However, early-type galaxies that reside in the blue cloud are found to be more unobscured in X--rays (-0.8\,$\lesssim$\,HR$(2-4.5keV/0.5-2keV)$\,$\lesssim$\,-0.5) compared with 
those belonging to the green valley (HR$(2-4.5keV/0.5-2keV)$\,$>$\,-0.5). This might suggest that part of the X--ray obscuration is due to the galaxy itself, moving the host-galaxy colours from blue cloud to 
red sequence. 
X--ray obscured early-type galaxies cover a wide range of hardness ratios (up to 1.0), with those sources belonging to the blue cloud and green valley having slightly lower obscuration 
compared to objects located in the red sequence. On the other hand, X--ray unobscured late-type sources seem to be more obscured than unobscured early-types, while for X--ray obscured 
late-types obscuration seems to be similar to that of obscured early-type galaxies. 

Recently, \cite{pierce10} studied the nuclear and outer U\,$-$\,B colours in relationship with X--ray obscuration, through the hardness ratio HR$(2-7keV/0.5-2keV)$, using a sample of 
X--ray selected AGN in the AEGIS field at redshifts 0.2\,$<$\,z\,$<$\,1.2. For most of their objects selected as X--ray type-1 AGN they obtained bluer colours, while for X--ray type-2 
sources hosts are characterised by redder colours. As shown in Figure~\ref{fig_cmd_nts}, no clear separation of such a type has been found between the B\,$-$\,V colour and the 
HR$(2-4.5keV/0.5-2keV)$ hardness ratio used in this work, when the complete population of AGN is observed. However, as described above, when we segregate X--ray obscuration for 
different morphological types, it seems that there might be a certain correlation between X--ray and optical colours.

\subsection{CMRs in relationship with X/O flux ratio}
\label{subsec_colour_xo}

Finally, we studied the CMRs in relationship with the X--ray-to-optical (X/O) flux ratio. The physical explanation of this parameter is still not evident, but as 
suggested in \cite{povic09b} it might be related with the accretion rate. Studying X--ray properties, such as the X/O flux ratio and X--ray luminosities, we can test the black hole growth rate in 
relationship with properties of host galaxies, colours and morphology.

\begin{figure*}[!ht]
\centering
\vspace*{1truecm}
\includegraphics[width=14.4cm,angle=0]{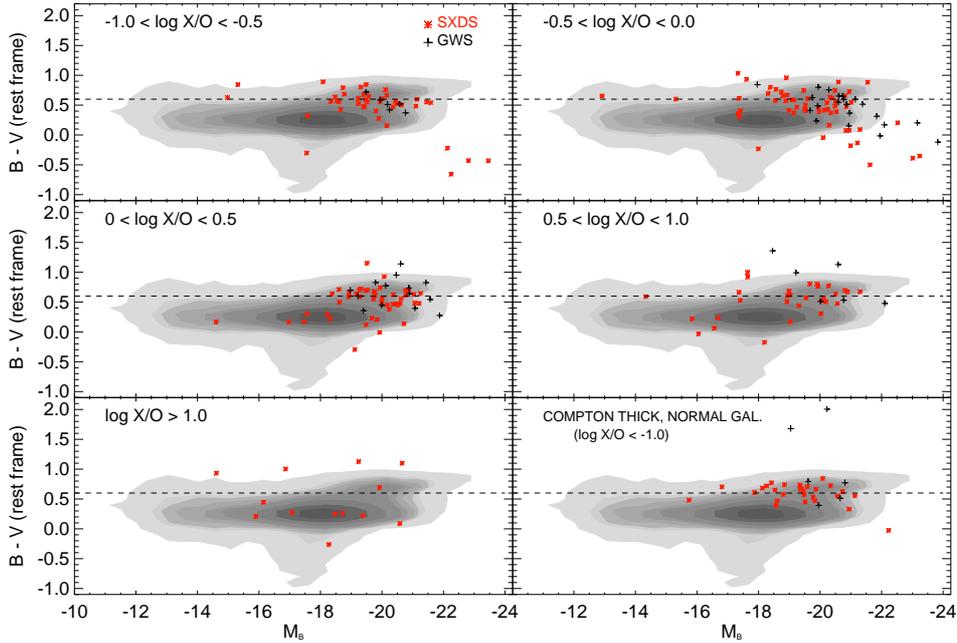}
\caption[ ]{Colour-magnitude diagram showing the relationship between the rest-frame B\,$-$\,V colour and the absolute magnitude in the B band for AGN in the SXDS (stars) and GWS (crosses) 
fields having different X--ray-to-optical flux ratio. All sources have redshifts z\,$\le$\,2.0. The sample of AGN is compared with the sample of normal galaxies in the CDF-S field 
\citep{wolf01,wolf04,wolf08} represented with contours. 
Grey scales of the contours are scaled to the data, where the darkest and brightest show the highest and the lowest density of the sources, respectively. 
The dashed line shows the \cite{melbourne07} limit between galaxies belonging to the red sequence (B\,$-$\,V\,$>$\,0.6) and those in the blue 
cloud (B\,$-$\,V\,$<$\,0.6). The bottom right diagram shows the distribution of objects classified as non AGN (Compton thick and normal galaxies), having very low values of the X/O flux ratio.
\label{fig_cmd_xo}}
\end{figure*}

Figure~\ref{fig_cmd_xo} shows the CMRs according to the X/O flux ratio. We have studied the distribution of AGN analysed in this work, considering five ranges of X/O flux ratio. 
We have also represented a distribution of objects having very low X/O flux ratios (logX/O\,$<$\,-1) characteristic of Compton thick sources and/or normal galaxies \citep[e.g.,][]{alexander01,fiore03,civano07}. 
As can be seen, most of these objects 
also reside in the green valley, having colours that correspond to most AGN sources. \\
\indent Examining Figure~\ref{fig_cmd_xo}, turns out that the distribution of AGN sources moves on the colour-magnitude diagram as the X/O flux ratio changes. Objects having 
higher values of the X/O flux ratio 
(e.g. logX/O\,$>$\,0.5) have bluer colours, while most AGN with lower X/O flux ratio have red and green colours with respect to that of normal galaxies. This might be related with 
two scenarios. 
First, that AGN with higher X/O flux ratio, have younger stellar populations, therefore showing bluer optical colours. In the case that the X/O flux ratio is a parameter proportional to the Eddington ratio, 
as suggested by \cite{povic09b}, and if larger accretion rates and stronger AGN activity provoke a stronger AGN feedback, then this might be in contrast to the usual prediction that the 
AGN feedback quenches the star formation, unless we are observing the initial phases when star-formation quenching is just beginning. And second, in the case of very high Eddington ratios, 
AGN do contribute to the optical colours of their host galaxies, making them bluer. We can also notice that all sources located in the region not covered by normal galaxies, 
being very blue, very luminous, and probably high redshift QSO sources, have lower values of X/O flux ratio ($\approx$\,logX/O\,$\le$\,0). Although the analysed sample of AGN seems to be representative of a 
full X--ray population as shown in Section~\ref{subsec_sample_selection}, we should be aware that the observed trends represent just a small fraction of the full X--ray population.

\begin{figure}[!ht]
\centering
\vspace*{1truecm}
\includegraphics[width=10.0cm,angle=0]{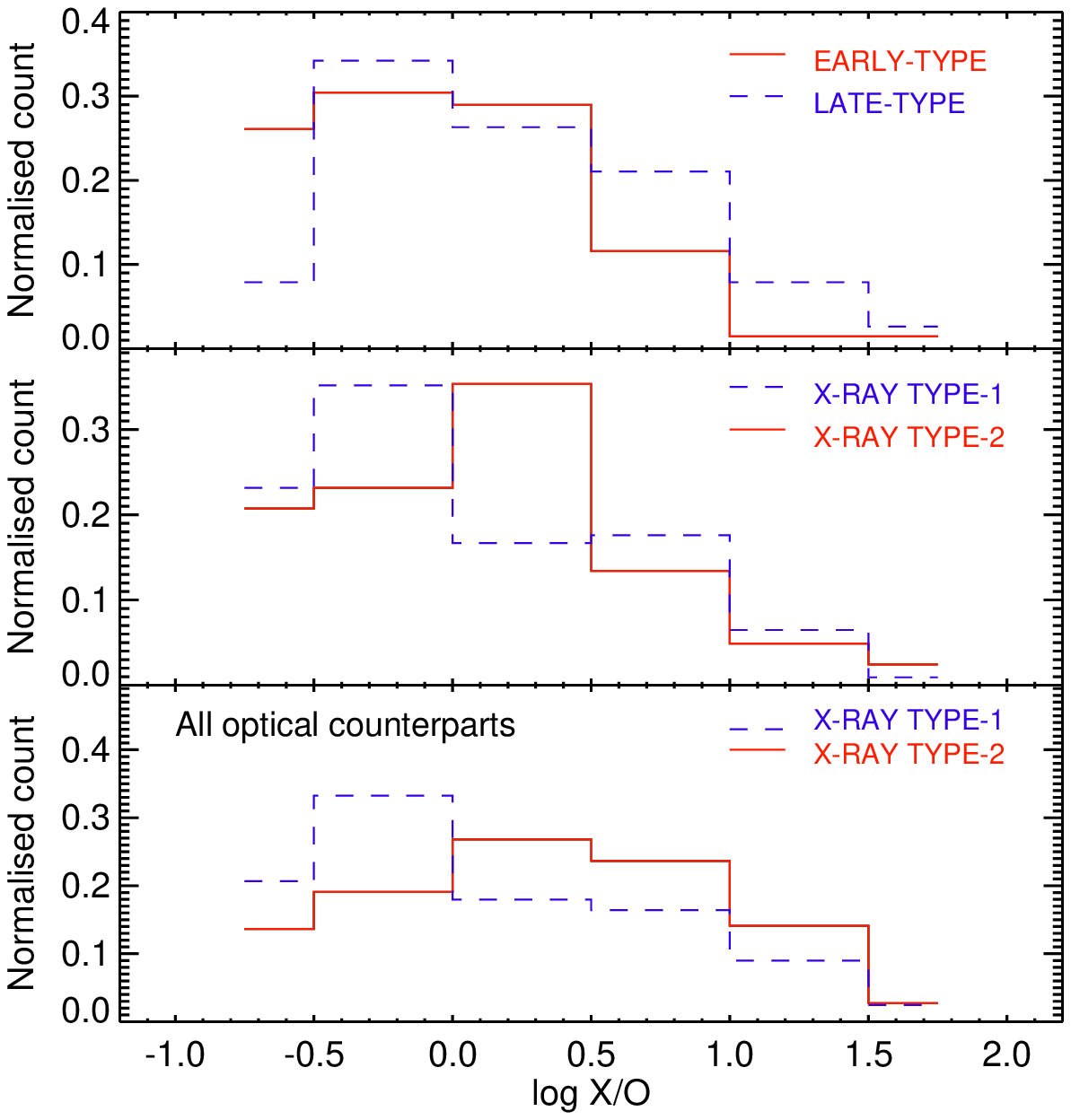}
\caption[ ]{Normalised X/O flux ratio distributions. (\textit{Top}) Two main morphological types, early- (solid red line) and late-type (dashed blue line) active galaxies, are presented. The other two panels 
represent X--ray type-1 (dashed blue line) and type-2 (solid red line) sources of a sample analysed in this paper (\textit{middle}) and of all optical counterparts (\textit{bottom}). 
\label{fig_histo_xo_morph_nucty}}
\end{figure}

\begin{figure}[!ht]
\centering
\vspace*{1truecm}
\includegraphics[width=8.0cm,angle=0]{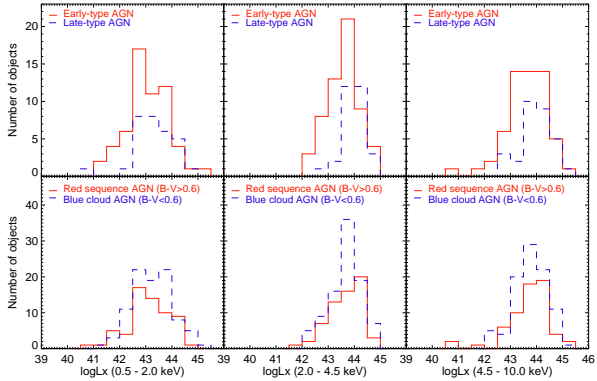}
\caption[ ]{\textit{From left to right}: Histogram showing a distribution of X--ray luminosities in three energy ranges: soft (0.5\,-\,2.0\,keV), hard-2 (2.0\,-\,4.5\,keV), and very hard (4.5\,-\,10.0\,keV), for 
\textit{Top} Early- (solid red line) and late-type (dashed blue line) active galaxies, and \textit{Bottom:} For red (solid red line) and blue (dashed blue line) AGN, having B\,$-$\,V colours $>$\,0.6 and $<$\,0.6, 
respectively.
\label{fig_histo_Lx}}
\end{figure}

Figure~\ref{fig_histo_xo_morph_nucty} shows the normalised distributions of X/O flux ratio for early- and late-type galaxies, and for X--ray unobscured and obscured sources of a sample analysed in 
this paper (see Section~\ref{subsec_sample_selection}) and of all optical counterparts. It can be seen that high values of the X/O flux ratio (X/O\,$>$\,0.5) correspond to AGN being hosted by later-types. This 
has already been obtained by \cite{povic09b}, suggesting that late-type AGN, having more material to feed the black hole, have higher Eddington ratios compared with early-type AGN. Recently, \cite{mainieri11} 
obtained the same result for Type-2 QSOs, where disk-dominated or merging systems in their sample have higher accreting rates in comparison with bulge-dominated galaxies. Moreover, we compared the X--ray luminosities in three 
energy ranges, soft, hard, and veryhard (see Table~\ref{tab_enrang_ecf_flux}) for early- and late-type active galaxies, as can be seen in Figure~\ref{fig_histo_Lx}. Performing the Kolmogorov-Smirnov statistic, it 
was found that the 
distributions are different in the three ranges, being significantly different in hard and veryhard X--rays with probability factors of only 0.0004 and 0.025 that they belong to the 
same parent distribution. Again, it can be seen that AGN hosted by later-types show higher X--ray activity compared with early-types. On the other hand, when we observe X--ray luminosities 
with respect to the B\,$-$\,V colour, dividing all sources into red (B\,$-$\,V\,$>$\,0.6) and blue (B\,$-$\,V\,$<$\,0.6) active galaxies, the Kolmogorov-Smirnov analysis suggests that 
their X--ray activities are not significantly different, having probabilities of 93\%, 68\% and 81\% in three energy ranges, respectively, and that they belong to the same parent distribution.

Finally, we compared the normalised distributions of X/O flux ratios for X--ray type-1 and type-2 AGN of a sample analysed in this paper (middle panel) and of all optical counterparts (bottom panel). We 
found similar distributions between two samples, where lower values of the X/O flux ratio ($<$\,0) are more populated by X--ray unobscured AGN.

\subsection{Summary: our results and current models of AGN formation and evolution}
\label{subsec_colour_summary}

\indent As shown in Figure~\ref{fig_cmd_allz}, and analysed above, AGN hosts seem to have different colours as compared to normal galaxies, with the peak of red/blue AGN host galaxies 
being moved to bluer/redder colours, respectively. As mentioned in Section~\ref{subsec_colour_morph}, in order to understand better what we see on the colour-magnitude diagrams and to relate the results 
obtained with the existing 
models of AGN formation and evolution, it is necessary to study colour-magnitude relations in more detail and instead of observing whole populations of active galaxies, to observe different 
morphological and nuclear types. As has been seen, almost all AGN classified as compact objects (possible QSO sources) are found to be X--ray unobscured and blue galaxies. However, when observing the two main morphological types, 
early- (which seem to be the majority of X--ray detected AGN) and late-type galaxies, no relationship has been found between colours and morphology, showing that early-/late-type X--ray 
selected AGN are not necessarily red/blue as non-active galaxies, respectively. In order to explain the presence of similar ('green') colours for both morphological types being located in the 
green valley, there might be two possibilities: \\
\indent 1) \textit{Early- and late-type X--ray detected AGN have \textbf{different stellar populations, similar to those of normal galaxies} belonging to the red sequence and blue cloud, respectively}. In this case, two 
mechanisms could be 
responsible for the changing colours of AGN: AGN contribution, affecting more early-types and making them bluer, and dust reddening, affecting most of all late-types and  making them redder. 
However, we still lack proof for this scenario. As shown in Section~\ref{subsec_colour_z}, AGN contribution to optical colours has shown to be negligible in most (if not all) 
previous analyses \citep{kauffman07,nandra07,silverman08,cardamone10}, for at least 90\% of AGN \citep{pierce10}. Moreover, we have seen that it is still not very clear how great the 
influence of the dust reddening effect could be on the optical colours \citep{cardamone10,xue10}. However, even if the dust-reddening effect is significant (moving later-types to bluer colours after extinction correction), 
since most AGN reside in early-type systems (being less affected by dust), after extinction correction most of AGN might still be residing in the green valley, without showing the 
colour bi-modality typical of non-active galaxies. \\
\indent 2) \textit{Early- and late-type X--ray detected AGN have \textbf{similar stellar populations, different to those of normal galaxies} belonging to the red sequence and blue cloud, respectively}. AGN 
might be hosted by similar galaxies, being later early- and earlier late-type sources, presenting one phase in the evolution of galaxies. Moreover, in the redshift intervals studied in this work 
(see Figure~\ref{fig_cmd_zinter}), we have seen that most green valley AGN are found at redshifts 0.5\,$\le$\,z\,$\le$\,1.5. This possibility could support the hypothesis 
that green valley AGN are transition objects, between the blue starburst galaxies and massive, red ellipticals, proposed already by various authors 
\citep[e.g.,][]{springel05,schawinski06,nandra07,georgakakis08,silverman08,treister09}.

\indent We independently analysed three regions in the diagrams depending on their absolute magnitudes, and compared the results with the current models of AGN formation and evolution. Taking into 
account that no clear trend has been found on the CMR's when observing different morphological, X--ray, or X/O types, we should be cautions with them when studying galaxy evolution. As we move from the right to the 
left on the colour-magnitude diagrams (from higher to lower B band absolute magnitudes), we may interpret what we see in CMRs 
(Figures~\ref{fig_cmd_allz}, \ref{fig_cmd_zinter}, \ref{fig_cmd_morph_allz}, \ref{fig_cmd_nts}, and \ref{fig_cmd_xo}) in the following way:\\
\indent - \textit{Region I:} Very high magnitudes (M$_B$\,$<$\,-21.0), very blue colours. This region corresponds to the one unpopulated by normal galaxies in our colour-magnitude diagrams. All AGN have high 
redshifts, from 1.5 to 2.0. They are all very blue, compact, and X--ray unobscured sources, having low values of hardness ratios. We may suspect that these sources are high-redshift quasars, 
and the AGN being in the so called 'QSO-mode' \citep{hopkins05a,hopkins05b,springel05,croton06,hopkins08a,hopkins08b}. In this mode, galaxy mergers and interactions are main mechanisms 
triggering nuclear activity in galaxies, providing enough material for both, star formation and supermassive black hole feeding. At the beginning the QSO is obscured by gas, but eventually 
it blows the obscuring and star-forming gas out from the galaxy through different AGN feedback mechanisms: ionisation, heating, radiation pressure \citep[e.g.,][]{ciotti07} and/or through strong 
winds and jets \citep[e.g.,][and references therein]{ciotti09,shin10}. After that, the AGN can become unobscured, and the light coming from the nuclear region could dominate the light coming from 
the host galaxy. \\
\indent - \textit{Region II:} From intermediate to high magnitudes (-21.0\,$<$\,M$_B$\,$<$\,-18.0), green colours. This is the region where $>$\,60\% of AGN from our sample are located, being mostly 
concentrated in the green valley, at the end 
of the blue cloud and at the bottom of the red sequence. As already mentioned and shown, the majority of X--ray detected AGN populate this region at redshifts 0.5\,$\le$\,z\,$\le$\,1.5, showing 
different levels of nuclear activity (wide range of X/O flux ratio). The region is populated by compact sources, being again mostly unobscured and blue, but also by early- and 
late-type active galaxies, although AGN hosted by early-types are more numerous, as shown in Section~\ref{sec_morph}. They have similar colours, and show similar levels of 
obscuration, being both unobscured and obscured in X--rays. It is possible that at these redshifts we are observing one of the peaks of nuclear activity in the evolution of galaxies 
\citep[e.g.,][]{hasinger05}, related with the peak of starburst activity, with a possible dynamical delay \citep{davies07,schawinski09, wild10,hopkins11}. In order to explain the 
properties of the AGN observed in this region and in the redshift range considered, it seems that we need different mechanisms responsible for AGN fuelling. They include major and minor 
mergers \citep[e.g.,][]{surace00,springel05,dimatteo05,hopkins05a,hopkins05b,cox08,hopkins08a,hopkins08b, somerville08}, but 
also some secular mechanisms, such as minor interactions, disk instabilities, nuclear and large scale bars, colliding clouds, and supernovae explosions 
\citep[e.g.,][]{kormendy04,wada04,hopkins09,cisternas11}. Thus, it is not strange that most of the AGN we are observing are hosted by early-type galaxies, being 
triggered through major/minor mergers. On the other hand, late-type AGN were probably triggered by some secular mechanisms, although recently it has been shown 
that in some cases discs can survive major mergers \citep{springel05b,robertson06,governato09}. Therefore, what we are observing in this region are probably again AGN 
in the 'QSO-mode' (objects classified as compact), and phases of AGN activity before and after the 'QSO-mode', that we see as obscured and unobscured early- 
and late-type AGN, respectively, triggered as suggested above. An additional possibility for triggering at these redshifts could be also the 'radio-mode' accretion \citep{croton06}, 
active after the bright QSO phase, as a result of static hot halo formed around the host galaxy and the accretion of this hot gas on to a supermassive black hole. However, this 
mechanism is usually related with lower luminosities, and low accretion rates.  \\
\indent - \textit{Region III:} Low optical magnitudes (M$_B$\,$>$\,-18.0), wide range of colours, from red to blue. Most AGN in this region are found at redshifts, z\,$\le$\,1.0. Around 30\% of sources are compact, being 
normally blue and unobscured, while the remaining objects have been classified as early- or late-type galaxies, both unobscured and obscured 
in X--rays. What we observe in this region could be, on one side post 'QSO-mode' AGN, that we see as unobscured early- and late-type AGN, triggered by some of the mechanisms described above, 
including AGN with the 'radio-mode' accretion, and on the other side new triggered AGN, at these redshifts probably by some secular mechanisms \citep[][and references therein]{cisternas11}.

\section{Conclusions}
\label{sec_conclusions}

The main objective of this work was to study the connection between the AGN and the host galaxy, in order to derive clues for understanding some of the still unanswered 
fundamental questions related with the AGN fuelling mechanisms, and their formation and evolution. To achieve the proposed objectives, we studied morphology and colours, two key 
elements for analysing the properties of the host galaxies, in relation with X--ray properties describing the AGN activity (X--ray luminosities, X--ray obscuration, X/O flux ratio). 
We chose the SXDS field, observed in X--rays with XMM-\textit {Newton}, which has one of the deepest optical data. After reducing X--ray data and after source detection, 
we derived a catalogue of 1121 X--ray emitters and cross-matched it with the publicly available optical catalogue. We obtained a catalogue of 808 X--ray emitters with optical counterparts, 
and for these sources we measured photometric redshifts and k--corrections, in order to obtain their rest-frame colours, absolute magnitudes, and luminosities. We obtained reliable photometric 
redshift information for 308 sources, for which we performed the analysis of their morphology and colours. We increased the SXDS sample by adding our 
previous data from the GWS fields.

The morphological classification still seems to be one of the main challenges, especially at high redshifts where obtaining reliable morphological information becomes very difficult. 
In this paper we used the galSVM \citep{huertas08,huertas09} code, one of the new methods for morphological classification, especially useful when dealing with high redshift sources, 
and with the possibility of using a range of different morphological parameters and non-linear boundaries to separate different types. Moreover, we used SExtractor \citep{bertin96} 
in order to obtain the input parameters for galSVM execution, and the CLASS\_STAR parameter, for separating between compact and extended sources. Using both galSVM and SExtractor, we obtained the 
following:

- A set of morphological parameters, including different parameters related with the concentration of the host galaxy light and its asymmetry.

- The final morphological classification, separating all AGN between compact sources and those hosted by one of the two main morphological types, namely 
early- and late-type galaxies. Approximately 22\% of our objects have been classified as compact sources, while $\approx$\,30\% of the AGN have been estimated to be hosted by early-type, 
and $\approx$\,18\% by late-type galaxies. For about 23\% of AGN, hosts were not very well resolved and the AGN might be residing in either early- or late-type galaxies, or experiencing possible 
interactions or mergers. Around 7\% of AGN remained unidentified.  

- Different problems, making morphological classification more difficult, have been recognised and tested in this paper. These include systematic trends of morphological 
parameters with source brightness, size, and redshift, low S/N ratios, and the number of parameters needed for the morphological classification. It has been seen that there is a trend 
of all observed morphological parameters with the apparent magnitude, size, and distance, and that parameters related with galaxy concentration (especially the M$_{20}$ parameter 
compared with concentration index and/or Gini parameter) instead of with asymmetry seem to be the most affected. Moreover, it has been seen that the combination of two or three parameters 
is not enough to have a reliable morphological classification, as has been done with all previous non-parametric methods, and that classification in the multi-parameter space, 
with simultaneous use of different parameters and non-linear boundaries is needed.

- At redshifts z\,$\le$\,2.0, at least 50\% of X--ray detected AGN analysed in this work are hosted by spheroids and/or bulge-dominated galaxies. However, at least 18\% of AGN in our sample are hosted by late-types, 
suggesting that different mechanisms can be responsible for triggering the nuclear activity in galaxies. \\

We studied colours of X--ray selected AGN through colour-magnitude relations, comparing the AGN distribution with the typical distribution of normal galaxies. First, we used all types of active galaxies to  analyse 
their redshift distributions (until z\,$\le$\,2.0). Second, 
using a high-redshift sample, we observed for CMRs in relation with morphology the first time, observing the distribution of active galaxies, belonging to different morphological types, on the colour-magnitude diagrams. 
Third, we studied the distribution of AGN on the colour-magnitude diagrams in relation with X--ray obscuration 
and, finally, in relation with X/O flux ratio. We conclude the following:

- Observing all types of X--ray detected AGN the highest number of sources is found to reside in the green valley, at the top of the blue cloud, and at the bottom of the red sequence, 
without showing any colour bi-modality typical for normal galaxies. AGN in our sample populate this region at redshifts $\approx$\,0.5\,--\,1.5. However, a higher number of low 
luminosity AGN have been detected in this work, due to the high depth of the optical SXDS data in comparison with the most (if not all) previous surveys. This allows us to study the colour and morphological 
properties of these objects and to compare them with the properties of high luminosity AGN. 

- More than 85\% of objects classified as compact (possible QSO sources), are found to have blue optical colours and to be unobscured in X--rays. However, when observing the two main 
morphological types, no correlation has been found between the colours and morphology. The AGN studied in this work do not show the standard trend of normal galaxies, where early-types, having older stellar 
populations, have redder colours, while late-type galaxies, being characterised with recent stellar formation have blue optical colours. AGN contribution and dust obscuration effects may affect the colours of early-/late-type 
galaxies moving them toward blue cloud/red sequence, respectively. However, additional work is still needed in order to quantify these contributions. On the other hand, even when mass-matched samples are used, no clear 
bi-modality has been seen in the optical colours. Moreover, both early- and late-type AGN are found to have similar ranges of X--ray obscuration, being both unobscured and obscured in X--rays.

Our findings confirm some previous suggestions that X--ray selected AGN residing in the green valley probably represent the transition population, quenching star formation and 
evolving to red sequence galaxies. What we might observe in the green valley is one of the peaks of AGN activity, with major and minor mergers probably being main triggering mechanisms, 
but also with the possibility of having some of the secular mechanisms responsible for fuelling. We observe AGN in the 'QSO-mode' (being compact, blue, and unobscured in X--rays) and different 
phases before (seeing them as obscured in X--rays) and after (seeing them as unobscured in X--rays) the 'QSO-mode', with AGN being hosted by later early- (the majority of sources) 
and earlier late-type galaxies, with similar stellar populations.

We would like to stress again that although a sample analysed in this paper seems to be representative of a full sample of X--ray population, all results presented here have been obtained for a sample of 25\% of 
the full population.

Finally, the paper provides the scientific community with a catalogue of a large sample of AGN with the X--ray and optical data obtained in this work, including final morphological classification, all derived morphological 
parameters, rest-frame colours, and photometric redshifts. The complete catalogue of all 1121 X--ray sources, 808 optical counterparts, and the objects analysed in this paper is available in the electronic version 
of this paper, while the description of columns and the small example for seven objects are presented in the Appendix.

\begin{acknowledgements}
We thank the anonymous referee for a detailed analysis of the paper and constructive comments. We also thank Isabel M\'arquez P\'erez, Josefa Masegosa Gallego, Jack Sulentic, and 
Ascensi\'on del Olmo Orozco for long and very useful discussions.\
This work was supported by the Spanish {\it Plan Nacional de Astronom\'\i a y Astrof\'\i sica} under grant AYA2011-29517-C03-01. JIGS acknowledges financial support 
from the Spanish Ministry of Science and Innovation under project AYA2008-06311-C02-02. MP acknowledges Junta de Andaluca and Spanish Ministry of
Science and Innovation through projects PO8-TIC-03531 and AYA2010-15169. We thank the SXDS, CDF-S, and COMBO-17 teams for making their data available to the astronomical community. We acknowledge support from 
the Faculty of the European Space Astronomy Centre (ESAC). We thank XMM-\textit{Newton} Helpdesk for their helpful comments during the X--ray data reduction.
This research has made use of software provided by the XMM-\textit{Newton} Science Operations Centre and Chandra X-ray Center (CXC) in the application packages SAS and CIAO, respectively. 
IRAF is distributed by the National Optical Astronomy Observatory, which is operated by the Association of Universities for Research in Astronomy (AURA) under cooperative agreement 
with the National Science Foundation. This publication makes use of data products from the Two Micron All Sky Survey, which is a joint project of the University of Massachusetts and the Infrared 
Processing and Analysis Center California Institute of Technology, funded by the National Aeronautics and Space Administration and the National Science Foundation.
\end{acknowledgements}

\begin{appendix}

\section{The catalogue}

The full catalogue of data obtained in this work will be available in the electronic edition of this paper. The catalogue contains SXDS X--ray data for all 1121 detected objects, optical identifications for the 
808 sources, and measured morphological parameters, final morphological 
classification, redshifts, and rest-frame colours for 308 X--ray emitters with optical counterparts described in Section~\ref{subsec_sample_selection}. 
Table~\ref{tab_cat_morph} shows an example of the format and content of the catalogue. The column entries are as follows:

Column 1 (ID): Identification number

Columns 2 and 3 (RA$_x$, DEC$_x$): Equinox J2000.0 right ascension and declination in degrees of the centroid in the X--ray catalogue. 

Column 4 (F$_s$): X--ray flux in the soft (0.5\,-\,2\,keV) band in 10$^{-15}$\,erg\,s$^{-1}$\,cm$^{-2}$

Column 5 (F$_h$): X--ray flux in the hard (2\,-\,4.5\,keV) band in 10$^{-15}$\,erg\,s$^{-1}$\,cm$^{-2}$

Column 6 (F$_{vh}$): X--ray flux in the hard (4.5\,-\,10\,keV) band in 10$^{-15}$\,erg\,s$^{-1}$\,cm$^{-2}$

Column 7 (F$_{tot}$): X--ray flux in the hard (0.5\,-\,10\,keV) band in 10$^{-15}$\,erg\,s$^{-1}$\,cm$^{-2}$

Column 8 (F$_{vh2}$): X--ray flux in the hard (4.0\,-\,7\,keV) band in 10$^{-15}$\,erg\,s$^{-1}$\,cm$^{-2}$

Column 9 (F$_{tot2}$): X--ray flux in the hard (0.5\,-\,7\,keV) band in 10$^{-15}$\,erg\,s$^{-1}$\,cm$^{-2}$

Column 10 (SXDS\_ID): Object name in the SXDS optical catalogue \citep{furusawa08}

Columns 11 and 12 (RA$_o$, DEC$_o$): Equinox J2000.0 right ascension and declination in degrees. These coordinates correspond to the centroid in the broadband optical catalogue obtained by the SXDS team 
(see Section~\ref{sec_data})

Columns 13 and 14 (R$_c$, R$_c$\_err): R$_c$ apparent magnitude and its error

Columns 15 and 16 (zphot, zphot\_err): photometric redshift and its error (see Section~\ref{subsec_zphot_kcorr}) 

Columns 17 and 18 (MabsB, MabsB\_err): Absolute magnitude in B band and its error (see Section~\ref{subsec_colour_z}) 

Columns 19 and 20 (B\,$-$\,V, B\,$-$\,V\_err): Rest-frame B\,$-$\,V colour and its error (see Section~\ref{subsec_colour_z}) 

Columns 21 and 22 (X/O, X/O\_err): X--ray-to-optical flux ratio, computed as the ratio of the observed X--ray flux in the 
0.5\,-\,4.5\,keV energy range and optical flux in R$_c$ band, and its error 

Columns 23 and 24 (HR, HR\_err): (2-4.5/0.5-2 keV) hardness ratio and its error (see Section~\ref{sec_data})  

Column 25 (Stellarity): SExtractor CLASS\_STAR parameter Column objects assigned as compact are all objects with this parameter $\ge$\,0.9 (see Section~\ref{sec_morph_method})

Column 26 (Elong): Elongation parameter obtained by SExtractor \citep{bertin96}

Column 27 (MSB): Mean Surface Brightness of the source, measured by galSVM \citep{huertas08}

Column 28 (A): Asymmetry index measured by galSVM, defined as in \cite{abraham96}

Column 29 (C): Abraham concentration index, measured by galSVM and defined as the ratio between the integrated flux within certain radius defined by the normalized radius α = 0.3, and the total flux \citep{abraham96}

Column 30 (Gini): Gini coefficient measured by galSVM and defined as in \cite{abraham03} 

Column 31 (S): Smoothness of the source, measured by galSVM and defined as in \cite{conselice03} 

Column 32 (M$_{20}$): Moment of light M$_{20}$, measured by galSVM and defined as in \cite{lotz04} 

Columns 33 and 34 (p1, p2): Probability that the galaxy belongs to early- or late-type, respectively (see Section~\ref{sec_morph_class})

\begin{table*}[!ht]
\begin{center}
\caption{Catalogue presenting morphological properties, colours, and photometric redshifts of X--ray emitters with optical counterparts in the SXDS field. 
In the electronic edition each object is represented per one line.
\label{tab_cat_morph}}
\small{
\begin{tabular}{c c c c c c c c c}
\noalign{\smallskip}
\hline\hline
\noalign{\smallskip}
\textbf{ID}&\textbf{RA$_x$}&\textbf{DEC$_x$}&\textbf{F$_s$}&\textbf{F$_h$}&\textbf{F$_{vh}$}&\textbf{F$_{tot}$}&\textbf{F$_{vh2}$}&\textbf{F$_{tot2}$}\\
\textbf{SXDS\_ID}&\textbf{RA$_o$}&\textbf{DEC$_o$}&\textbf{R$_c$}&\textbf{R$_c$\_err}&\textbf{zphot}&\textbf{zphot\_err}&\textbf{MabsB}&\textbf{MabsB\_err}\\
\textbf{B\,$-$\,V}&\textbf{B\,$-$\,V\_err}&\textbf{X/O}&\textbf{X/O\_err}&\textbf{HR}&\textbf{HR\_err}&\textbf{Stellarity}&\textbf{Elong}&\textbf{MSB}\\
\textbf{A}&\textbf{C}&\textbf{Gini}&\textbf{S}&\textbf{M$_{20}$}&\textbf{p$_1$}&\textbf{p$_2$}\\
\hline

1&34.2059326&-4.9195004&0.3803&2.959&5.493&2.842&11.63&3.617\\
SXDS-iC-121387&34.20587083&-4.918922&23.745&0.012&1.89 &0.09&-20.840&0.0111\\
0.077&0.021&0.635 &0.316& 0.295&0.561&0.03&1.349&26.994\\
0.098&0.450&0.690&0.039 &-1.293&0.167&0.833\\
&&&&&\\	
2&34.2367325&-4.7958641&5.128&18.22&17.888&20.502&38.996& 25.817\\
SXDS-iC-1616&34.23629167&-4.796003&22.373&0.004&0.615&0.004&-19.264&0.003\\
0.701 &0.0150&1.255 &0.198&-0.089&0.131&0.04&1.178&24.827\\
0.072&0.488&0.757&0.052 &-1.964&0.703&0.297\\
&&&&&\\	
3&34.3014336&-4.8216581&1.458&2.807&5.575&5.763&12.206&7.190\\
SXDS-iC-157690&34.30122083&-4.821953&24.704&0.021&1.02 &0.01& -19.291&0.009\\
0.588 &0.047&1.960  &0.904 &-0.163&0.406&0.58&1.829&26.736\\
0.093&0.449&0.719&0.063&-1.664&0.908&0.092\\
&&&&&\\	
4&34.3149796&-5.0099578&2.910&2.602&2.188&6.941&0.0&8.819\\
SXDS-iC-086715&34.31469167&-5.010572&22.232&0.003&1.42 &0.01&-22.251  &0.005\\
-0.657&0.009&0.260 &0.077&-0.499&0.228&0.97&1.071&25.493\\
0.120&0.538&0.822&-0.0      &-1.523&0.989&0.011\\
&&&&&\\	
5&34.3232346&-4.9310899&1.083&4.228&2.751&3.707& 8.398&5.0123\\
SXDS-iC-115045&34.3228583 &-4.931603&21.780&0.003&0.845&0.007&-20.484&0.003\\
0.5301  &0.007&0.165&0.0544&-0.041&0.292&0.03&1.325&25.331\\
-0.005&	0.501&0.760&-0.153&-1.919&0.711&0.289\\	
\noalign{\smallskip}
\hline
\end{tabular}
}
\end{center}
\end{table*}

\end{appendix}

\end{document}